\documentclass[final,3p,times]{elsarticle}

\journal{High Energy Density Physics}

\begin{document}

\begin{frontmatter}

\title{Simulating radiative shocks in nozzle shock tubes}

\author[label1]{B. van der Holst\corref{cor1}}
\ead{bartvand@umich.edu}
\author[label1]{G. T\'oth}
\author[label1]{I.V. Sokolov}
\author[label1]{L.K.S. Daldorff}
\author[label2]{K.G. Powell}
\author[label1]{R.P. Drake}

\address[label1]{Department of Atmospheric, Oceanic and Space Sciences,
  University of Michigan, Ann Arbor, MI 48109, USA}
\address[label2]{Department of Aerospace Engineering,
  University of Michigan, Ann Arbor, MI 48109, USA}

\cortext[cor1]{Corresponding author}
\begin{abstract}

We use the recently developed Center for Radiative Shock Hydrodynamics (CRASH)
code to numerically simulate laser-driven radiative shock experiments. These
shocks are launched by an ablated beryllium disk and are driven down
xenon-filled plastic tubes. The simulations are initialized by the
two-dimensional version of the Lagrangian Hyades code which is used to evaluate
the laser energy deposition during the first $1.1\,$ns. The later times are
calculated with the CRASH code. This code solves for the multi-material
hydrodynamics with separate electron and ion temperatures on an Eulerian
block-adaptive-mesh and includes a multi-group flux-limited radiation
diffusion and electron thermal heat conduction. The goal of the present paper
is to demonstrate the capability to simulate radiative shocks of essentially
three-dimensional experimental configurations, such as circular and elliptical
nozzles. We show that the compound shock structure of the primary and wall
shock is captured and verify that the shock properties are consistent with
order-of-magnitude estimates. The produced synthetic radiographs can be used
for comparison with future nozzle experiments at high-energy-density laser
facilities.

\end{abstract}


\end{frontmatter}

\section{Introduction}

In the experiments of the CRASH project, high-energy-density plasma flow is
driven by a beryllium disk irradiated by the Omega laser beams.
The ablation of part of the beryllium results in a rocket launch of the
remaining beryllium driving a strong shock-wave through a xenon-filled plastic
tube. The shock heats the xenon gas to sufficiently high temperature so that
it will ionize and create free electrons. Directly behind the shock the
electrons and ions will equilibrate due to the Coulomb collisions. This
will heat the electrons. The shock is fast enough so that the electrons have
to emit radiation behind the shock in order to satisfy the energy balance
equation. This results in a radiative cooling layer \cite{reighard2007}. The
transport of the radiation ahead of the shock will heat and ionize the xenon
in the radiative precursor. Some fraction of the radiation in the upstream
region will also transport sideways and strike the plastic tube. These photons
can heat the plastic ahead of the primary shock. The ablated plastic that
moves then inward compresses the xenon resulting in a so-called wall shock
\cite{doss2009,doss2011}.
It is the interplay between the primary and wall shock that is of interest
for the CRASH project.

To model the laser-plasma physics and laser energy deposition in the radiative
shock tube experiments, we use H2D, the 2D version of the Hyades code
\cite{larsen1994}, which contains a built-in laser package. H2D is a
Lagrangian radiation-hydrodynamics code that utilizes the axisymmetry. Hyades
is capable of tracing rays in 3D; the runs shown below used 2D ray tracing for
the laser energy deposition.
We use the H2D code to simulate our experiments for the first $1.1\,$ns
(sometimes up to $1.3\,$ns), the time of the full width half maximum (FWHM)
$1\,$ns laser pulse including
ramp-up and ramp-down time. The spatial profile of the beam is determined
using a model for the irradiance pattern of the laser beams in typical
experiments. At $1.1\,$ns the experiment is in a regime
that is well described by radiation-hydrodynamics so that the simulation can
be continued with the CRASH code \cite{vanderholst2011} instead. This code
solves for the multi-material hydrodynamic equations with a multi-group
flux-limited diffusion model for the radiation and uses the recently developed
Block Adapative Tree Library (BATL) \cite{toth2011}. We are
currently constructing a laser package in our simulation code as an
alternative to H2D and the progress in that work will be reported elsewhere.

The main goal of the CRASH project is to assess and to improve the
predictive capability of a simulation code, based on the combination of
experiments, simulation studies and statistical analysis. The baseline
experiment is a laser-driven shock launched through a xenon-filled straight
tube. We plan to perform an experiment in which the tube geometry is
changed to contain a circular nozzle. These experiments and
the comparison with the simulations will demonstrate how well the numerical
code can predict the shock properties of the following experiment in which a
wide circular tube tapers into an elliptical tube. In this project we plan to
analyze the radiative shock structures in nozzle shock tubes in the context of
predictive capability and uncertainty quantification of the simulations
\cite{holloway2011}. The demonstration that such numerical simulations are
feasible is the topic of the present paper.

This paper is organized as follows. Section \ref{sec:circular} describes
the radiative shocks produced in nozzles with circular cross-section. We
contrast the results with order-of-magnitude estimates based on physical
arguments. Section \ref{sec:elliptical} demonstrates the radiative shocks
in nozzles with an elliptical neck. This simulation is fully three-dimensional.
The paper concludes with Section \ref{sec:conclusions}.

\section{Circular nozzle}\label{sec:circular}

A laser pulse irradiates a $20\,\mu$m thick beryllium disk
with $0.35\,\mu$m wavelength light for the FWHM
duration of $1\,$ns and with a laser energy deposition of $4\,$kJ. For the
laser spot size we use a FWHM $800\,\mu$m diameter which is smaller than the
$1200\,\mu$m diameter of the tube. We are currently constructing a laser
package in the CRASH code, but the runs reported here use the Lagrangian
radiation-hydrodynamics code Hyades 2D (H2D) \cite{larsen1994} to evaluate the
laser energy
deposition during the first $1.1\,$ns which includes the laser ramp-up and
ramp-down time. While our aim is to simulate radiative shocks in nozzles, the
first $1.1\,$ns is however simulated in a straight tube for convenience.
Calculating the first $1.1\,$ns with a straight tube and then
transforming this tube into a nozzle is physically justified as long as
the taper and shaft of the nozzle does only alter the radiative precursor.
The justification originates from the observation that the radiative
transport from the precursor back through the shock front is negligible
\cite{drake2011}. The considered straight tube does have a cylindrical
polyimide wall of $100\,\mu$m thickness and inner radius of $600\,\mu$m filled
with xenon with mass density $\rho=0.0065\,$g/cm$^3$, while there is vacuum
outside. The beryllium disk is immediately to the left of $x=0$, where $x$ is
the coordinate along the tube. The laser light will come in from the negative
$x$ direction. We place a gold washer next to the beryllium disk to protect
the outside of the plastic tube from the laser light and use acrylic in
between the gold and polyimide tube.

\begin{figure}
\begin{center}
{\resizebox{0.48\textwidth}{!}{\includegraphics[clip=]{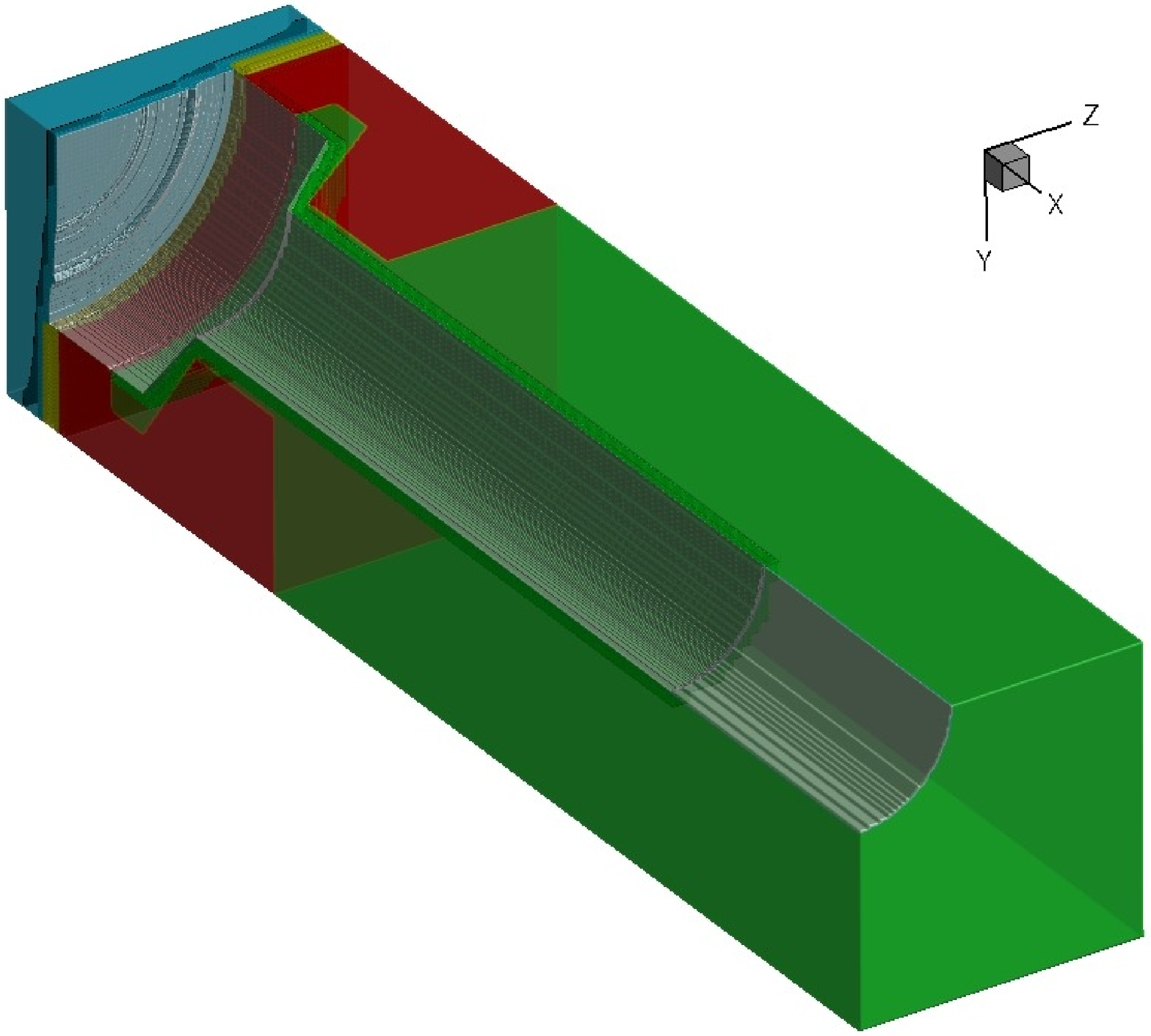}}}
{\resizebox{0.48\textwidth}{!}{\includegraphics[clip=]{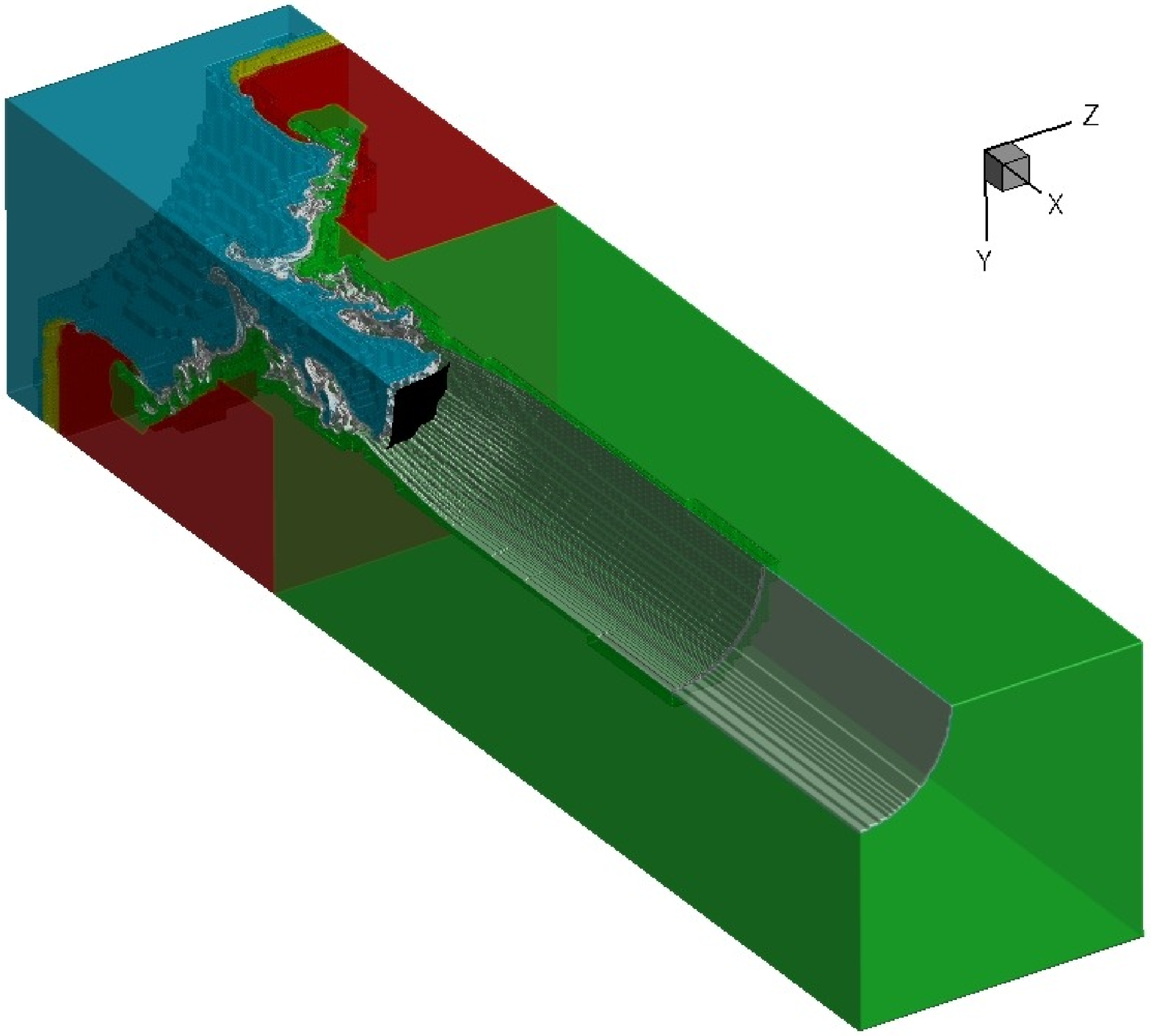}}}
\end{center}
\caption{The materials in the 3D circular nozzle experiment at $1.1\,$ns
(left panel) and $13\,$ns (right panels). Only one quarter of the shock tube
is shown. The materials are beryllium (blue), polyimide (green), acrylic (red)
and gold (yellow). The xenon inside the tube is not colored, but the border of
the xenon volume is indicated by the white surface. The primary shock at
$13\,$ns is depicted by a black iso-surface.}
\label{fig:circular_3D}
\end{figure}

After $1.1\,$ns of simulation time the output of H2D is used to initialize the
Eulerian CRASH code as described in \ref{sec:init}. The straight tube of H2D is
transformed into a nozzle, see the left panel of Fig. \ref{fig:circular_3D}
for one quarter of the nozzle domain.
This nozzle changes cross-section in the following way. For $x<500\,\mu$m
we do not modify the tube as defined in H2D. For $x>750\,\mu$m we shrink the
tube diameter from $1200\,\mu$m to $600\,\mu$m and the polyimide wall thickness
corresponding reduces to $50\,\mu$m. Between $500\,\mu$m and $750\,\mu$m the
tube diameter and wall thickness are linearly shrinking (In the notation of
\ref{sec:init} we use $x_0=500\,\mu$m, $x_1=750\,\mu$m,
$\varepsilon_y=1/2$ and $\varepsilon_z=1/2$). The vacuum outside the
nozzle is replaced with low density polyimide to avoid the otherwise zero
mass density. The left panel of Fig. \ref{fig:circular_3D} shows the materials
that are present in the simulation in color: beryllium (blue), polyimide
(green), acrylic (red), and gold (yellow). The xenon inside the nozzle is for
convenience not shown in this figure.

The computational domain size is $-150<x<3900$, $0<y<900$ and
$0<z<900$ in microns, where $y$ and $z$ are the two directions transverse to
the nozzle. The simulation is performed with an effective resolution of
$2560\times 512\times 512$ grid cells using two levels of refinement.
The effective cell sizes are therefore approximately $1.6\,\mu$m along the
tube and $1.8\,\mu$m in the two transverse directions. The
domain is decomposed in $4\times 4\times 4$ grid blocks. The mesh is refined
at all interfaces that involve xenon or gold. To capture the shock front and
the cooling layer, the mesh is also refined where the xenon density exceeds
$0.02\,$g/cm$^3$. Grid blocks are also refined if these criteria are satisfied
in the ghost cells.

We perform the simulations with
the CRASH radiation hydrodynamics code \cite{vanderholst2011}. This code
solves for the radiation hydrodynamic equations in three operator splitting
steps: (1) an explicit time step of the hydrodynamic equations using a
shock-capturing solver, (2) a linear advection of the radiation bins in
frequency-logarithm space, and (3) an implicit solve of the radiation
diffusion, heat conduction, and energy exchanges. This code is continuously
undergoing improvements. One such improvement is the resolution change
treatment for the radiation diffusion and electron heat conduction, see
\ref{sec:reschange}. During the simulations of the nozzles these improvements
were not yet present and the original scheme in \cite{vanderholst2011} was
used instead. We use for the hydrodynamic part of the equations the HLLE
scheme with a Courant-Friedrichs-Lewy number of 0.8 and the generalized Koren
limiter with $\beta=3/2$. For the radiation, we use
the multi-group flux-limited diffusion model with 30 groups.
The photon energy range is $0.1\,$eV to $20\,$keV that is logarithmically
distributed over the groups. The radiation diffusion, heat conduction and
energy exchanges are solved with the split (decoupled) implicit solver
of \cite{vanderholst2011} using the conjugate gradient method with a
Schwarz-type Incomplete Upper-Lower (ILU) preconditioner.

Due to the symmetry in the problem we only need to simulate one quarter of the
nozzle domain ($y>0$ and $z>0$) and use reflective boundary conditions at
$y=0$ and $z=0$. For all other boundaries of the domain we use extrapolation
with zero gradient. For the radiation we use zero albedo boundary conditions.

We simulated the shock evolution from $1.1\,$ns to $13\,$ns physical time. It
took a little over three days to compute on 1000 cores of the HERA
supercomputer at the Lawrence Livermore National Laboratory. The number of
cells in the computational domain increased from $26.5$ million at the
beginning to about 38 million near the end. The 3D material identity at
$13\,$ns is shown in the right panel of Fig. \ref{fig:circular_3D}.
The beryllium has moved into the nozzle and is like a piston driving a shock
in the xenon. This shock is indicated by a black iso-surface (of
high ion temperature values). The xenon itself is not shown but the edge of the
xenon is emphasized by a white color to make the xenon entrainment between
the beryllium and polyimide more clear. We can also see the inward moving
polyimide which will lead to a wall shock. In the following we will analyze
the shock structure in more detail and check if the results are in agreement
with back-of-the-envelope estimates presented in \cite{drake2006,drake2011}.

\begin{figure}
\begin{center}
{\resizebox{0.48\textwidth}{!}{\includegraphics[clip=]{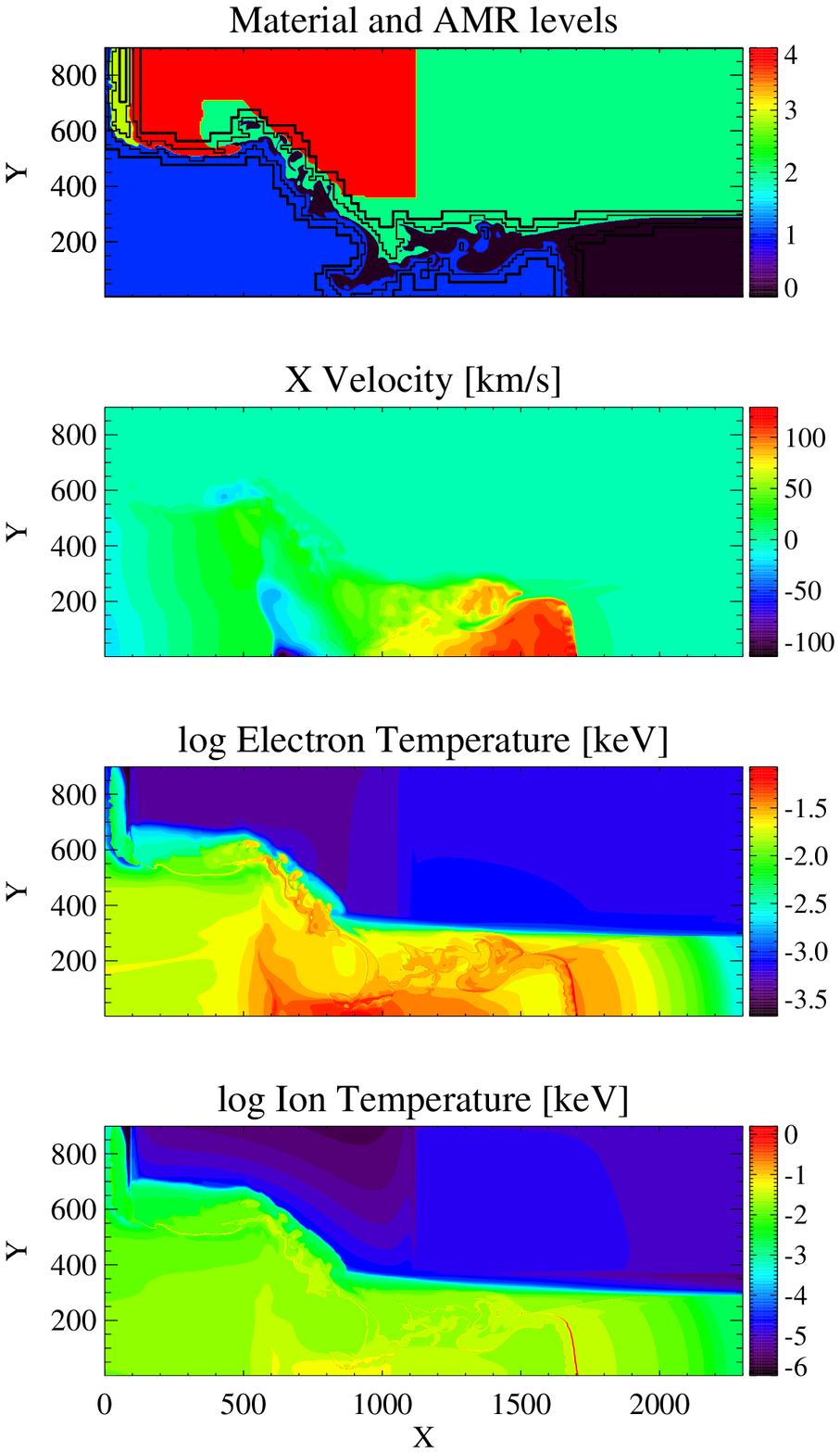}}}
{\resizebox{0.48\textwidth}{!}{\includegraphics[clip=]{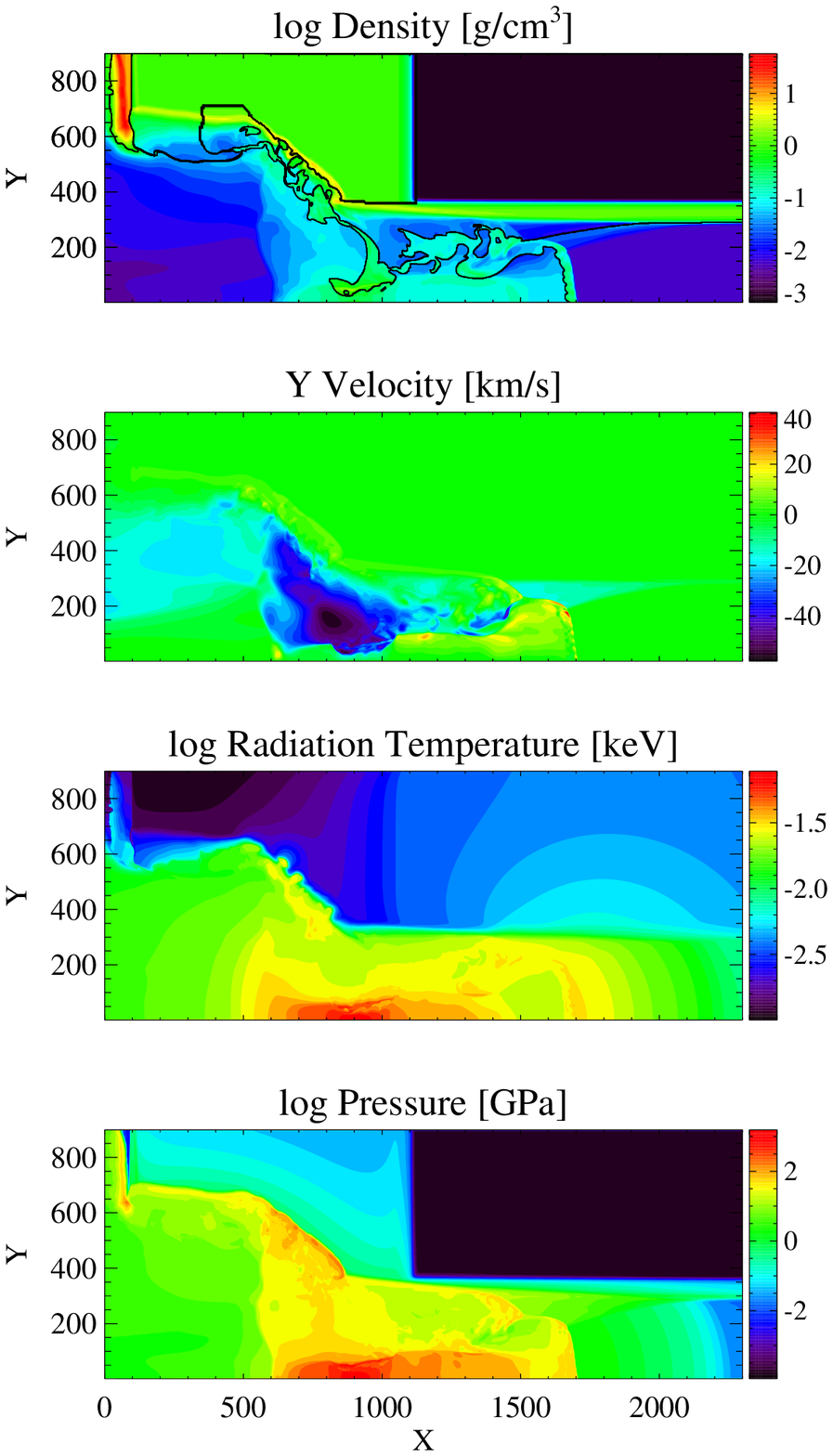}}}
\end{center}
\caption{The radiative shock structure in the 3D circular nozzle simulation
at $13\,$ns. The plots show in color contour the plasma and radiation state
indicated in the plot titles as a function of the $x$ and $y$ positions in
microns. The colors in the top left panel indicate beryllium (blue),
xenon (black), polyimide (green), gold (yellow) and acryllic (red). The black
lines in this panel show resolution changes, while in the top right panel the
lines indicate the material interfaces.}
\label{fig:circular_state}
\end{figure}

The shock structure in the $xy$-plane at time $13\,$ns is shown in Fig.
\ref{fig:circular_state}. The top left panel is for
the material identification of beryllium (blue), xenon (black), polyimide
(green), gold (yellow) and acrylic (red). The nozzle with inner radius of
$600\,\mu$m, taper, and shaft with inner radius of $300\,\mu$m are visible.
The black lines indicate the resolution changes between the grid refinement
levels. The top right panel shows the mass
density. Part of the polyimide is of very low density and represents vacuum.
The dense polyimide tube thickness ranges from $50$ to $100\,\mu$m. The
xenon is compressed by the beryllium piston flow resulting in a
primary shock that is located at $x\approx 1700\,\mu$m. For convenience we have
indicated with black lines where the material interfaces are.

To check the obtained properties of the primary shock we first determine
the beryllium piston velocity. For a $1\,$ns laser pulse of $4\,$kJ
energy, the irradiance on a $800\,\mu$m diameter spot size is
$8\times 10^{14}\,$W/cm$^2$. Most of this light will be absorbed in the
beryllium, so that the absorbed energy per unit of area during this $1\,$ns is
$8\times 10^{5}\,$J/cm$^2$. About $20\%$ of the beryllium mass, corresponding
to $4\,\mu$m of the $20\,\mu$m, will be ablated by the laser \cite{drake2011}.
The ablation efficiency, which is the ratio of the kinetic energy of the
remaining $16\,\mu$m beryllium to the total kinetic energy and exhaust, is
to lowest order in the ablation percentage equal to the fraction of the mass
that initially has been ablated, i.e. the efficiency is $20\%$. In reality,
however, about half of the incident laser light reaches the beryllium
above the absorption region where ablation occurs, while the other half is
absorbed below the critical density. The actual efficiency from converting
laser energy to kinetic energy of the remaining $16\,\mu$m of beryllium is
therefore roughly $10\%$.  The areal mass density of $16\,\mu$m beryllium at
$1.8\,$g/cm$^3$ is approximately $3\times 10^{-3}$g/cm$^2$. The areal kinetic
energy density $0.1\times 8\times 10^{5}$J/cm$^2$ corresponds therefore to an
initial velocity of the beryllium of a little more than $v=200\,$km/s. This
beryllium will launch a shock through the xenon-filled tube. This high
velocity is only achieved at early times. The numerically obtained shock
velocity gradually reduces to a value between  $110\,$km/s and $120\,$km/s
at time $13\,$ns as shown in the X velocity plot of Fig.
\ref{fig:circular_state}.

\begin{figure}
{\resizebox{0.48\textwidth}{!}{\includegraphics[clip=]{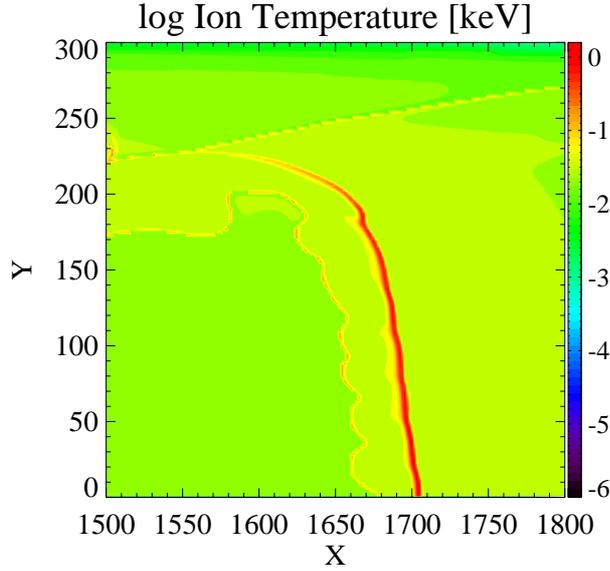}}}
\caption{Zoom-in of the ion temperature near the shock with the $x$ and $y$
coordinates in microns.}
\label{fig:circular_zoom}
\end{figure}

The unshocked xenon gas pressure is initially about $1.1\,$atm while the
density is $\rho=6.5\times 10^{-3}\,$g/cm$^3$. With the above mentioned shock
velocity of $v\approx 110\,$km/s we obtain a xenon post-shock pressure
at $13\,$ns of the order of $\rho v^2\approx 80\,$GPa. The post-shock pressure
at $x\approx 1700$ obtained in the bottom right panel of Fig.
\ref{fig:circular_state} is in agreement with this estimate. The shock wave
heats the ions. The ion temperature in the postshock region of
a strong shock wave with compression ratio equal to
$\kappa=(\gamma+1)/(\gamma-1)$ is for our application approximately
\cite{drake2006}:
\begin{equation}
  k_B T_{i} = Am_p v^2\frac{1-1/\kappa}{\kappa},
\end{equation}
where $k_B$, $m_p$ and $A$ are the Boltzmann constant, the proton mass and
atomic mass, respectively. For xenon with atomic mass $A=131$, a shock
velocity of $110\,$km/s, and an adiabatic index of $\gamma=5/3$ for ions
gives a postshock temperature of $T_i \approx 3\,$keV. We find
with the numerical simulation an ion temperature of $1.2\,$keV in Fig.
\ref{fig:circular_zoom} that shows the region around the shock. With an
effective resolution of $1.6\,\mu$m we do not fully resolve this narrow ion
temperature peak, resulting in a lower than expected maximum temperature. We
find two other temperature peaks as well. One peak is near $x=1600\,\mu$m
behind the shock and another is near $y=250\,\mu$m. They overlap with the
beryllium-xenon and the polyimide-xenon material interfaces, respectively.
These spikes are probably the result of a low-order convergence rate in the
interface treatment.

\begin{figure}
{\resizebox{0.48\textwidth}{!}{\includegraphics[clip=]{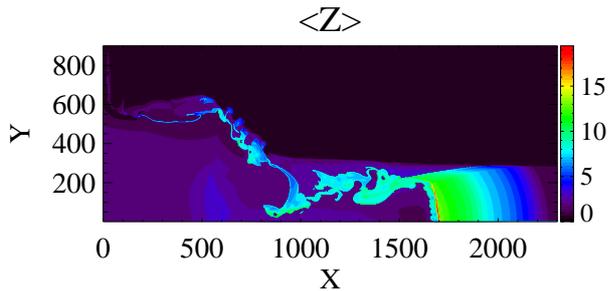}}}
\caption{The mean ionization in the circular nozzle simulation as a function
of the $x$ and $y$ coordinates in microns.}
\label{fig:circular_zavg}
\end{figure}

For a high density plasma the collision frequency between electrons and
ions is large, so that their temperature will equilibrate. At the shock,
however, the ion temperature jumps so that the ions and
electrons are out of equilibrium. The ions will heat the electrons via
Coulomb collisions and form an equilibration zone directly behind the shock.
The electron heating will also increase the ionization in the equilibration
zone. Fig. \ref{fig:circular_zavg} shows the average ionization in the
$xy$-plane. We find from the code that the ionization is elevated to about
$\left< Z \right> =17$ at the electron temperature peak in the equilibration
zone. At this peak the energy of each ion is shared with $\left< Z \right>$
electrons. The equilibration temperature $T_{\rm eq}$ can be approximated as
the postshock temperature for a strong shock under the assumption that
the electrons and ions are in temperature equilibrium:
\begin{equation}
  k_B T_{\rm eq} = \frac{1}{\left< Z \right> + 1}Am_p v^2\frac{1-1/\kappa}{\kappa}.
\end{equation}
For representative values of the polytropic index between $\gamma=1.2$ and
$1.3$ for single-temperature xenon, the estimated equilibration
temperature is between $T_{\rm eq} = 76\,$eV and $104\,$eV. In reality this
temperature will be somewhat lower due to radiative cooling. The numerically
obtained value is $T_e \approx 73\,$eV in Fig. \ref{fig:circular_state}.
This electron temperature is not the final state in the postshock region since
$\sigma T_e^4 \approx 2.8\times10^{12}\,$W/cm$^2$, where $\sigma$ is the
Stefan--Boltzmann constant. The incoming kinetic energy flux is however
$\rho v^3/2\approx 4.3\times 10^{11}\,$W/cm$^2$
based on the velocity at time $t=13\,$ns of $v=110\,$km/s and xenon density
$\rho=0.0065\,$g/cm$^3$. There is therefore not enough incoming energy.
There must be a cooling layer \cite{reighard2007} through which the electron
temperature falls to the final temperature $T_f$ estimated by
$2\sigma T_f^4 = \rho v^3/2 \approx 4.3\times 10^{11}\,$W/cm$^2$. The factor 2
is because the radiative cooling layer emits in both directions equally. The
final temperature is thus $T_f \approx 38\,$eV. 

The heated electrons are the main energy source for radiation. In Fig.
\ref{fig:circular_state} we show the radiation temperature as a measure for the
total radiation energy density. The photons travel upstream of the
shock where they preheat and ionize the unshocked xenon in the
precursor as depicted in the electron temperature panel in Fig.
\ref{fig:circular_state} and Fig. \ref{fig:circular_zavg}. The radiation
transport in the unshocked xenon is not diffusive and we rely on the
flux-limited diffusion to recover the optically thin free-streaming limit.
This free-streaming approximation is accurate enough as long as the radiation
transport from the precursor back to the shock is negligible, in contrast with
the almost omnidirectional photon distribution function as assumed in the
diffusive limit. A fraction of the upstream
radiation expands sideways and heats the polyimide wall ahead of the
primary shock. This will ablate the polyimide of the wall. The resulting
inward polyimide flow is visible in the Y velocity panel of Fig.
\ref{fig:circular_state}. This infow extends at $13\,$ns to
$x\approx 2100\,\mu$m and does have a magnitude of about $10\,$km/s. The exact
magnitude of this flow might depend on the radiation transport
fidelity used in simulations. The ablated polyimide compresses the xenon
as can be seen in the density panel in Fig. \ref{fig:circular_state} by the
faint tilted feature between $x\approx 1700\,\mu$m and $x\approx 2100\,\mu$m.
The resulting wall shock has an angle with the primary shock and their
shock properties were analyzed in \cite{doss2009,doss2011}.

The material identity in the top left panel of Fig. \ref{fig:circular_state}
also demonstrates the entrainment of xenon in between the beryllium and
polyimide. In \cite{doss2011} the entrained flow was shown to first get
shocked by the wall shock and then again shocked near the tripple point of
the wall shock and the primary shock. We also mention that in our simulations
the entrained shear flows produce Kelvin--Helmholtz roll-ups at for instance
$x\approx 1400\,\mu$m.

\begin{figure}
{\resizebox{0.48\textwidth}{!}{\includegraphics[clip=]{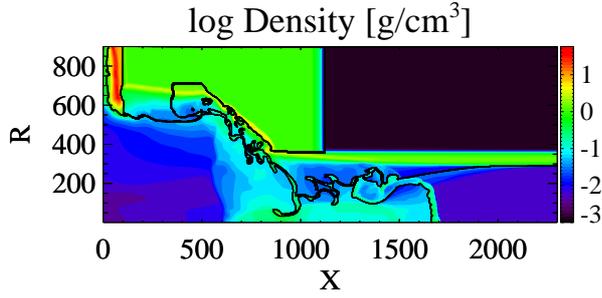}}}
\caption{The density at $13\,$ns in color contour. This simulation is performed
in the $rz$-geometry. The axes are in microns and the black line indicates the
change in material level.}
\label{fig:circular_2D}
\end{figure}

The simulation presented in this section was performed in 3D Cartesian
geometry. The axi-symmetry in the problem does however allow to perform this
simulation in 2D cylindrical $rz$-geometry as well. We used
the same settings for the numerical radiation-hydrodynamics solvers.
The computational domain is $-150<z<3900$ and $0<r<900$ in microns, where
$r$ is the radial coordinate and $z$ is now the coordinate along the tube.
The effective resolution is $2560\times 512$ using two levels of refinement,
so that the cell sizes in the $rz$-geometry correspond to the cell sizes in
the $xy$-plane in the 3D Cartesian simulation. The 2D mesh is decomposed
in $4\times 4$ grid blocks. The mesh refinement criteria are also the same as
for the 3D case. In Fig. \ref{fig:circular_2D} the density is shown at the
final time $13\,$ns. The results are quite similar to those of the previous
simulation. The main difference is in the entrained xenon, which requires
probably higher resolution to be fully resolved.

\begin{figure}
{\resizebox{0.48\textwidth}{!}{\includegraphics[clip=]{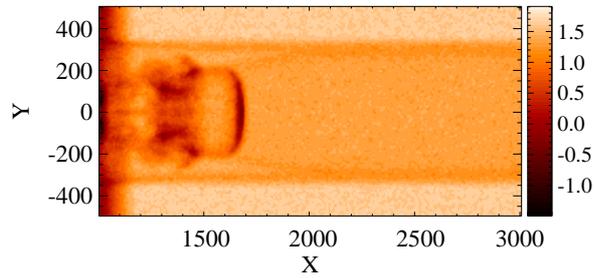}}}
\begin{center}
\end{center}
\caption{Synthetic radiograph image at $13\,$ns for the circular nozzle
simulation. The X-ray source is located at $(x,y,z)=(2,-12,0)$ in mm.}
\label{fig:circular_radiograph}
\end{figure}

The aim of the CRASH project is to predict certain properties of the compound
shock in our high-energy-density laser experiments. From the experiment we
obtain X-ray radiograph images using He-$\alpha$ emission from a backlit
pinhole source transmitting through the experimental target. These images
show fundamentally where the dense xenon is. To be able to make a direct
comparison between the observations and simulations, we added the capability
to the CRASH code to create synthetic radiograph images. The details of
the implementation and verification are presented in \ref{sec:radiograph}.
The radiograph for the circular nozzle simulation is shown in Fig.
\ref{fig:circular_radiograph}. We locate the X-ray source at $(x,y,z) = 
(2000,-12000,0)\,\mu$m. We have blurred the image to account for
the finite pinhole size and finite exposure time in the experiment. We also
added Poisson noise to mimic the finite photon count. The dense xenon behind
the primary shock at $x=1700\,\mu$m and the wall shock is clearly visible.
We can also see the entrained xenon and the Kelvin-Helmholtz roll-up to the
left of the primary shock.

\section{Elliptical nozzle}\label{sec:elliptical}

The setup of the elliptical nozzle is very similar to that of the circular
nozzle. The main difference is that the wide tube of $1200\,\mu$m inner radius
changes the cross-section down the tube into an ellipse with a major axis of
$1200\,\mu$m and a minor axis of $600\,\mu$m. The first $1.1\,$ns of the
simulation is, as for the circular nozzle, performed with the H2D
radiation-hydrodynamics code to determine the laser energy deposition.
For the tube geometry in H2D a straight tube is used with a diameter
of $1200\,\mu$m. The output of H2D is used to initialize the CRASH code
following the recipe outlined in \ref{sec:init}. If we use the notation
that $x$ is the coordinate along the tube and $y$ and $z$ are the two
directions transverse to the tube, then we remap the straight tube of H2D
to CRASH using the coordinate transformations shown by the equations
\ref{eq:nozzle_y} and \ref{eq:nozzle_z} in which $x_0=500\,\mu$m,
$x_1=750\,\mu$m,
$\varepsilon_y=1/2$ and $\varepsilon_z=1$. That means that the circular tube
is tapered into an elliptical shaft between $x_0=500\,\mu$m and
$x_1=750\,\mu$m. The domain size, effective resolution, refinement criteria,
boundary conditions, and the used numerical scheme are the same as for the
circular nozzle.

\begin{figure}
\begin{center}
{\resizebox{0.48\textwidth}{!}{\includegraphics[clip=]{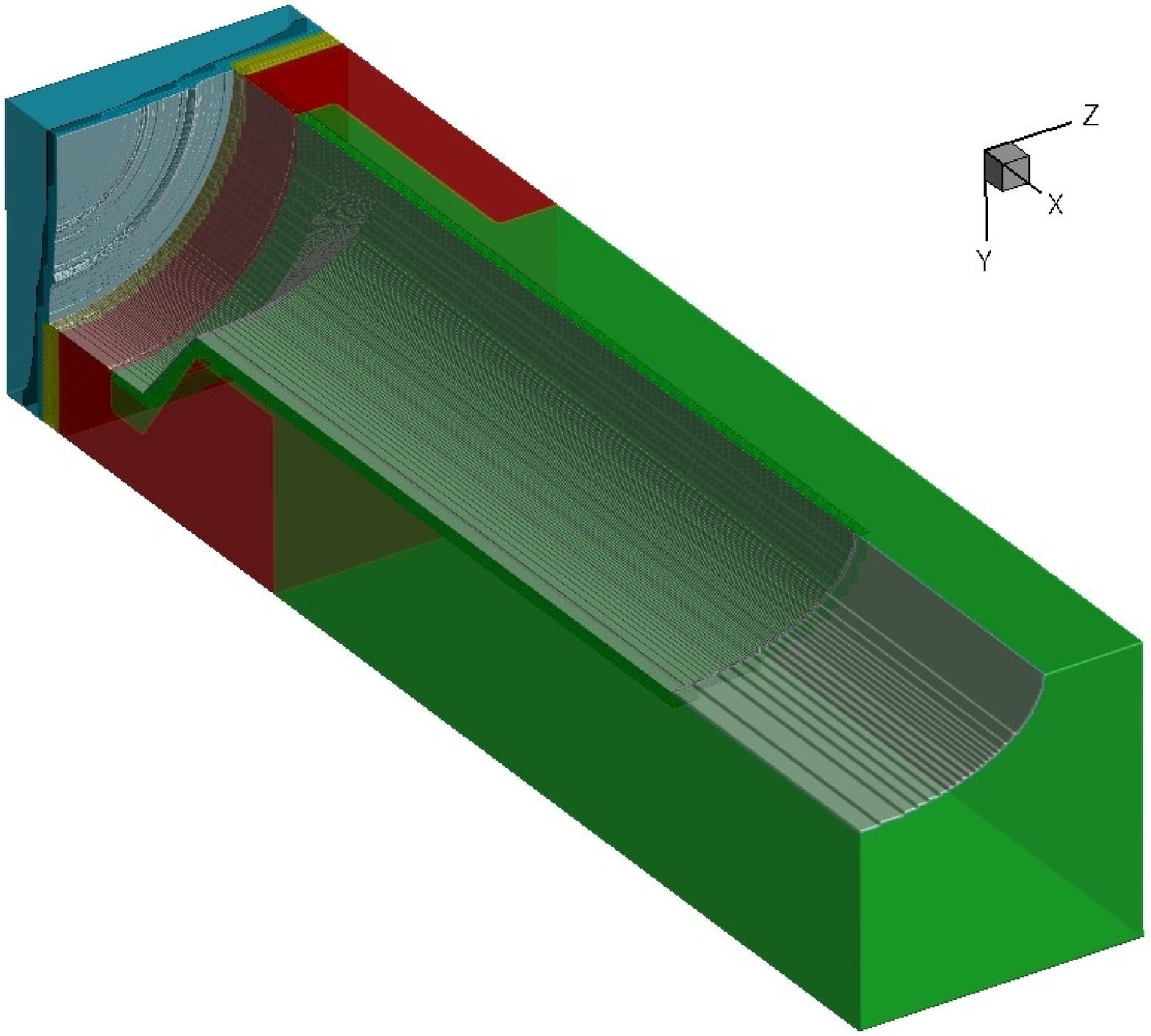}}}
{\resizebox{0.48\textwidth}{!}{\includegraphics[clip=]{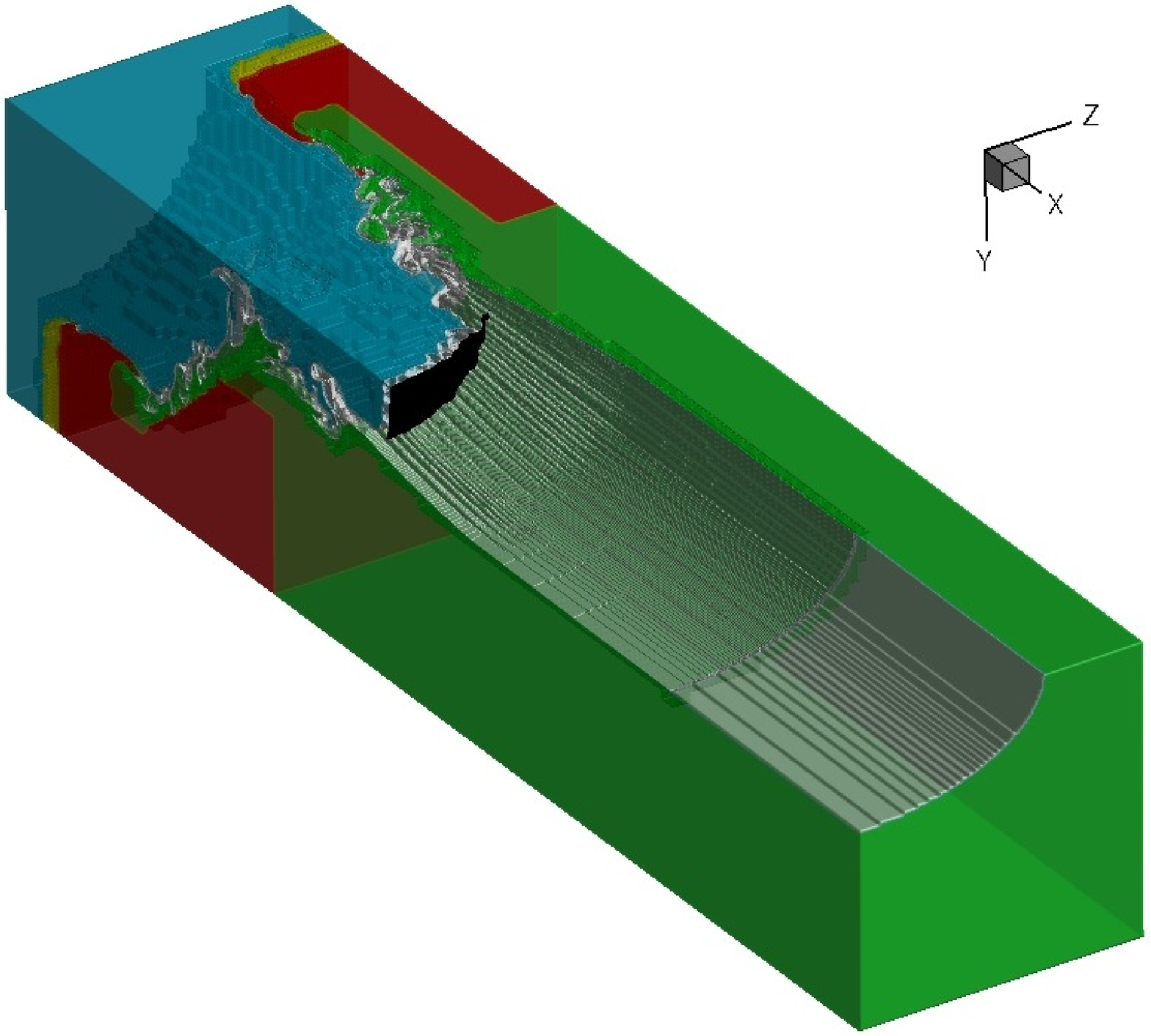}}}
\end{center}
\caption{The materials in the 3D elliptical nozzle simulation at $1.1\,$ns
(left panel) and $13\,$ns (right panel). One quarter of the nozzle is shown.
The color code for the materials is the same as in \ref{fig:circular_3D}. The
black iso-surface indicate the primary shock.}
\label{fig:elliptical_3D}
\end{figure}

The simulation is performed from $1.1\,$ns to $13\,$ns physical time. The
number of finite volume cells increased from $29.6$ million initially to
about $45.3$ million at the end of the simulation. The computational time
was $3.5$ days on 1000 processor cores of the HERA supercomputer. The 3D
material identification is shown in Fig. \ref{fig:elliptical_3D}. The left
panel is for $1.1\,$ns and the right panel for $13\,$ns. The color code is
blue for beryllium, green for polyimide, red for acrylic and yellow for gold.
Xenon is for convenience not colored in these panels so that the elliptical
shaft is visible. At $13\,$ns the beryllium has moved into this shaft and
drives a shock in the xenon like a piston. The shock front is indicated with a
black surface. The edge of the volume occupied by the xenon is colored in
white to visualize the entrainment of xenon between the polyimide and beyllium.

\begin{figure}
\begin{center}
{\resizebox{0.48\textwidth}{!}{\includegraphics[clip=]{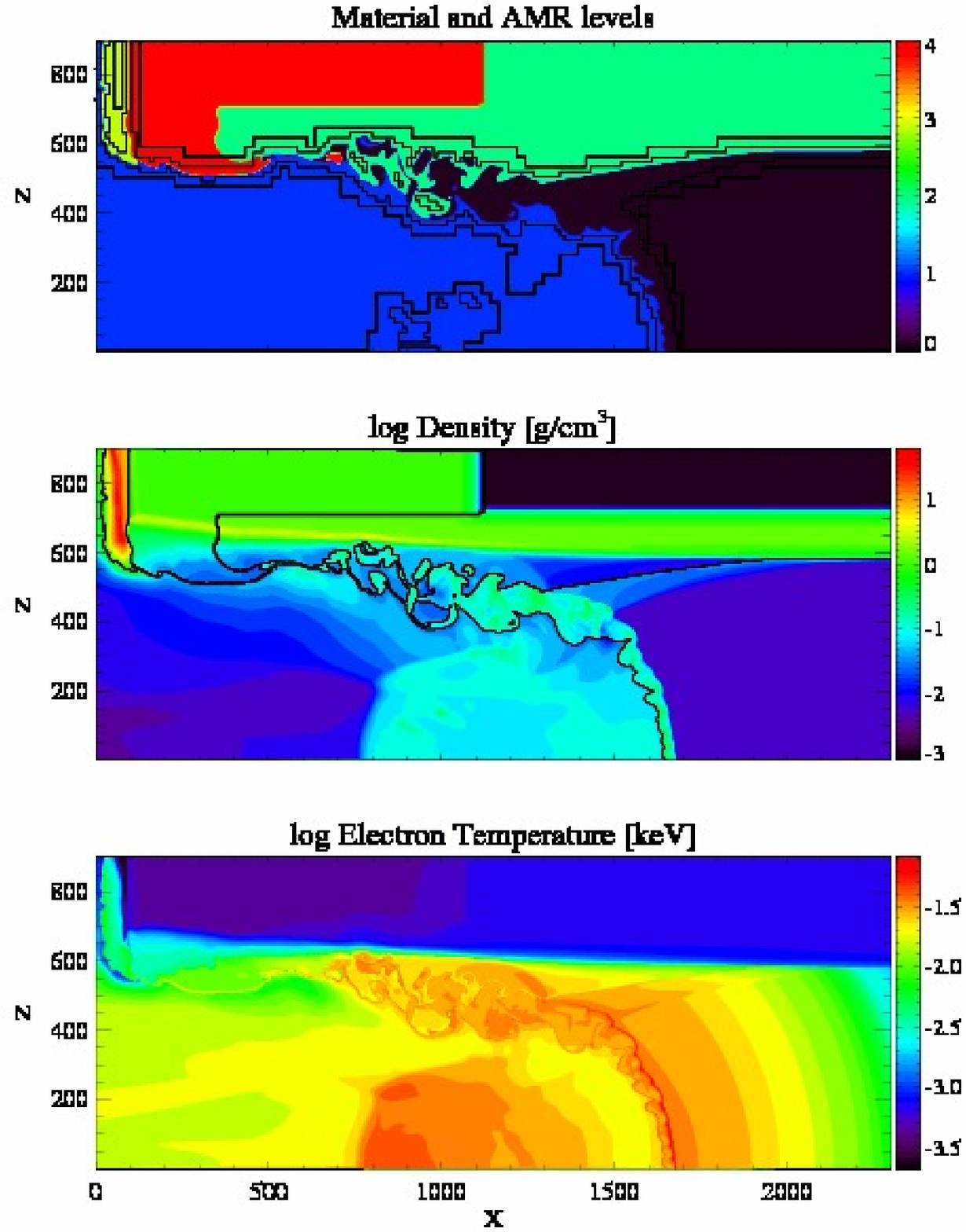}}}
{\resizebox{0.48\textwidth}{!}{\includegraphics[clip=]{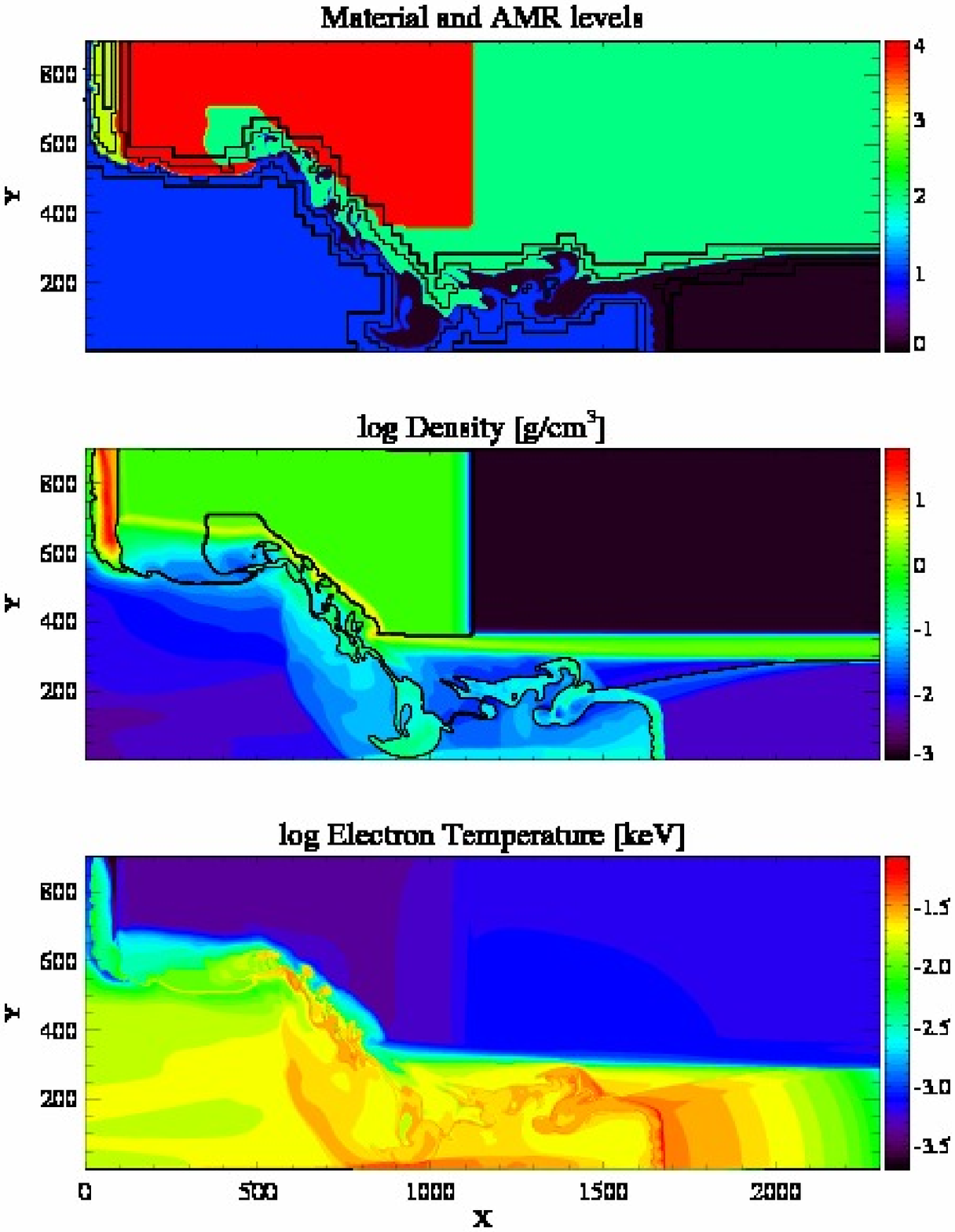}}}
\end{center}
\caption{The output of the 3D elliptical nozzle simulation at $13\,$ns. The
state variables indicated in the plot titles are shown in color as a function
of the $(x,z)$ coordinates given in microns in the $y=0$ plane on the left and
$(x,y)$ coordinates in the $z=0$ plane on the right. The color bars provide the
scales. The color code of the material levels in the top panels indicates
the same materials as in Fig. \ref{fig:circular_state}. The black lines in
the top panels are for the resolution changes, while the black lines in the
middle panels indicate the change in material indentity.}
\label{fig:elliptical_state}
\end{figure}

The left panels of Fig. \ref{fig:elliptical_state} show the material location,
mass density and electron temperature, respectively, in the $xz$-plane at
$13\,$ns. The basic ingredients of the compound
radiative shock structure are seen in these panels. The primary shock near
$x\approx 1700\,\mu$m in this plane is curved, since
the diameter of the laser spot of $800\,\mu$m is smaller than the major axis,
$1200\,\mu$m, of the elliptical shaft. The ripple in the compressed xenon
region behind the primary shock was analyzed for similar experimental
conditions \cite{doss2011b}. We also find again the tilted wall shock in front
of the primary shock. The electron temperature panel shows the temperature
peak in the equilibration zone behind the primary shock and a decreasing
temperature behind that in the radiative cooling zone. The right panels of
Fig. \ref{fig:elliptical_state} show the same physical quantities in the
vertical $yz$-plane. The shock structure in this plane is quite similar to the
results found for the circular nozzle.

\begin{figure}
\begin{center}
{\resizebox{\textwidth}{!}{\includegraphics[clip=]{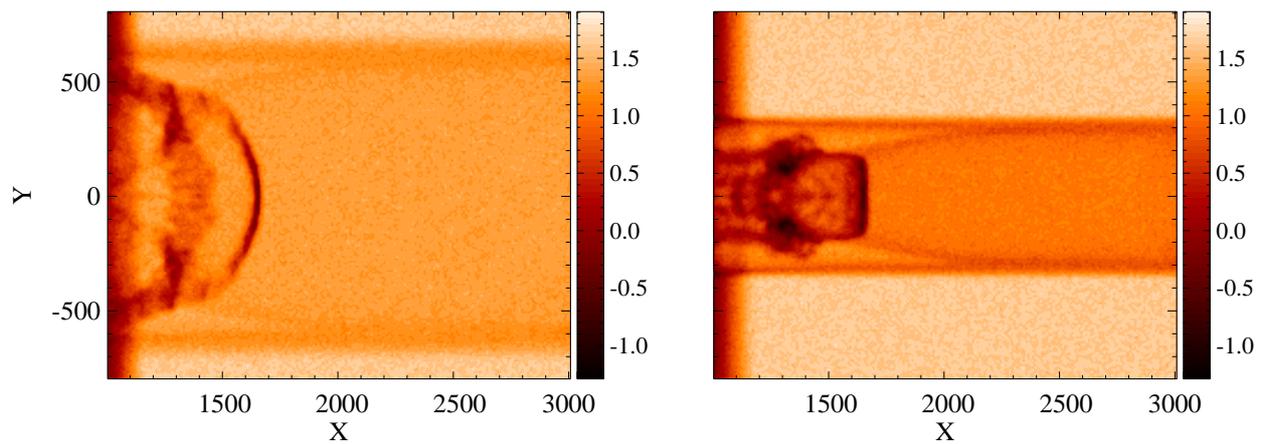}}}
\end{center}
\caption{Synthetic radiographs created along the two transverse axes of the 3D
elliptic nozzle at $13\,$ns. The X-ray sources are at $(x,y,z)=(2,-12,0)\,$mm
directed at the $y=0$ (left panel) and at $(x,y,z)=(2,0,12)\,$mm directed at
the $z=0$ plane (right panel).}
\label{fig:elliptic_radiograph}
\end{figure}

The difference in the shock structure and compressed xenon region behind the
shock should also be visible in the synthetic radiographs. Indeed, in Fig.
\ref{fig:elliptic_radiograph} we show in the left panel the image
produced by an X-ray source at $(x,y,z)=(2000,-12000,0)\,\mu$m directed at
the $xz$-plane and in the right panel the image produced by an X-ray source
at $(x,y,z)=(2000,0,12000)\,\mu$m directed at the $xy$-plane. The compressed
xenon behind the primary and wall shock are found as dark features in these
images. Note that the image in the right panel is darker than the image in the
left panel, since the rays are going through twice as much xenon (the major
axis is twice the minor axis). Also note that the rippled structure of the
compressed xenon layer behind the primary shock is somewhat smoothed out in
these images.

\section{Conclusions}\label{sec:conclusions}

In this paper we have discussed the simulations of radiative shocks in nozzle
shaped tubes. We recover basic properties found in laser-driven shock
experiments in high-energy-density facilities and from radiation hydrodynamic
theories: A compressed layer shows up directly behind the primary shock in
which the electrons are heated in an equilibration zone by the shock wave
heated ions. This is followed by a radiative cooling layer. The emitted
photons propagate upstream and preheat the precursor ahead of the primary
shock. The lateral expansion of this radiation ablates the plastic nozzle which
results in a wall shock. The simulations are shown to be in agreement with
simple back-of-the-envelope estimates.

We have used the two-dimensional version of the Hyades code to determine the
laser energy deposition during the first $1.1\,$ns of the experiment. This
code simulates on a Lagrangian mesh the radiation hydrodynamics based on a
multi-group flux-limited diffusion model and flux-limited electron thermal
heat conduction. We are currently constructing a laser package in the Eulerian
CRASH code so that the laser energy deposition and the produced radiative
shocks in the xenon at later times are self-consistently evaluated with the
same code.

We plan to perform laser-driven shock experiments in three-dimensional
circular and elliptical nozzles. The comparison of the elliptical nozzle
simulations with such experiments will stress test the performance of the
code. For the validation we can compare the properties of the shock structure
and compressed xenon layers from both the simulations and experiments.
These 3D properties can in the experiments be extracted from dual, orthogonal
radiography \cite{kuranz2006} and then compared to the two orthogonal synthetic
radiographs from the simulations. 

\section*{Acknowledgments}
This work was funded by the Predictive Sciences
Academic Alliances Program in DOE/NNSA-ASC via grant DEFC52-08NA28616 and by
the University of Michigan. The authors acknowledge M.J. Grosskopf and
E. Rutter for providing Hyades simulations. The authors would also like to
thank B. Fryxell and E. Myra for helping to test the CRASH code.

\appendix

\section{Initializing CRASH with Hyades}\label{sec:init}

The laser energy deposition during the first $1.1\,$ns is evaluated with
the Lagrangian radiation hydrodynamics code Hyades 2D (H2D) \cite{larsen1994}.
The output of H2D is on a distorted logically Cartesian mesh of
$n_x \times n_y$ cells. When we initialize the CRASH code with H2D, this
mesh is first triangulated by splitting each quadrilateral along the shorter
diagonal. The interpolation from the triangulated Hyades grid requires finding
the triangle that surrounds the center of a given grid cell of the CRASH code.
A simple linear search can become very inefficient when we have many grid
cells (order of a hundred thousand) per processor. To accelerate this method,
first we create a uniform grid with about 200 by 200 resolution that covers
the whole domain. For each rectangular cell in the uniform grid we find and
store the list of triangles that intersect it. This can be done very fast.
Then we perform the interpolation onto the CRASH grid by first finding the
rectangular cell that surrounds it, and then we only check the triangles that
intersect this cell to see which one contains the CRASH grid cell center.

The H2D simulations are performed on a Lagrangian grid where all cells
correspond to a unique material. These materials are identified by a material
index in the H2D output. Our code uses an Eulerian grid and we track the
material by means of level set functions. These level set functions are smooth
and signed distance functions, initialized
with the following algorithm. For each cell $i$ of the H2D grid having
material index $\alpha_i$, the level set function for material $\alpha_i$
is set to the distance to the closest H2D cell that contains a material
different from $\alpha_i$:
\begin{equation}
  d_{\alpha_i}(i) = + \min_{j,\alpha_j \neq \alpha_i} |{\bf r}_i - {\bf r}_j|,
\end{equation}
where ${\bf r}_i$ is the location of cell $i$. The level set functions for
the other materials, $\beta \neq \alpha_i$, are set to the negative distance
to the closest cell containing material $\beta$:
\begin{equation}
  d_\beta(i) = - \min_{j,\alpha_j= \beta} |{\bf r}_i - {\bf r}_j|.
\end{equation}
We interpolate these level set functions to the cells on the CRASH grid.
In our first-order level set scheme the material in the cell is indicated by
the largest level set function. For the simulations with H2D of our
shock tube experiments we use xenon, beryllium,
polyimide, gold, acrylic and ``vacuum''. In CRASH, vacuum is reassigned as
polyimide with low mass density. At later times during the evolution, the
location of material $m$ follows the simple advection equation for the level
set function
\begin{equation}
  \frac{\partial d_m}{\partial t} + \nabla\cdot(d_m{\bf u})
  = d_m\nabla\cdot{\bf u}.
\end{equation}
Also here, the material having the largest $d_m$ is assigned to be the material
of the cell.

A 3D nozzle is created from 2D Hyades as follows. We first
read the H2D shock tube output and triangulate the data. After that we
transform the coordinates in the transverse directions $y$ and $z$. The
circular cross section of the tube shrinks into an ellipse or smaller circle
by means of
\begin{eqnarray}
  y' &=& y\left[1 + (\varepsilon_y-1)
    \max\left(0,\min(1,\frac{x-x_0}{x_1-x_0})\right)\right],
  \label{eq:nozzle_y}\\
  z' &=& z\left[1 + (\varepsilon_z-1)
    \max\left(0,\min(1,\frac{x-x_0}{x_1-x_0})\right)\right],
  \label{eq:nozzle_z}
\end{eqnarray}
where $\varepsilon_y$ and $\varepsilon_z$ are the factors with which the $y$
and $z$ coordinates contracts for $x>x_1$ along the tube. This shrinking
varies linearly between $x_0$ and $x_1$. For $x<x_0$ the tube is not modified.
This means that the plasma at the far end is relocated closer towards the axis.
This will change the output of H2D to an unphysical state (the solution of the
straight tube is not the same as the solution of the nozzle). The impact of
this change in geometry is however minimal since at $1.1\,$ns the shock
dynamics did not yet reach the far end of the shock tube.

\section{Synthetic radiographs}\label{sec:radiograph}

The CRASH code has the capability of generating synthetic radiographs during
the simulations. The time frequency of the plotting, the location (or possible
locations) of the X-ray source(s) and the orientation, size, and number of
pixels of the radiograph images are all input parameters.

In the CRASH shock tube experiments we typically produce radiographs by
transmitting $5.18\,$keV
(mainly V He-alpha line) X-rays through material whose temperature is of the
order of $50\,$eV. These X-rays interact only weakly with the free or outer
electrons. We can therefore use cold opacities for the materials in the
experiment to evaluate the transmission. X-ray photons at this energy
experience negligible refraction in the experimental target. So it is
legitimate for our radiograph calculation to use straight-line analysis.

These images are line-of-sight plots that calculate the optical depth along the
straight lines (rays), connecting the X-ray source to the center of the image
pixel. The optical depth is an integral of the mass density multiplied by the
specific opacity characteristic for the material and the spectrum of the X-ray
source along the ray:
\begin{equation}
  D = \int_0^\infty \rho({\bf r}(l)) \kappa({\bf r}(l)) dl,
  \label{depth}
\end{equation}
where $\kappa$ is the specific opacity of the materials used in the
experiments and the ray is parameterized by the distance $l$ to the source.
In our shock-tube experiments \cite{drake2011},
we assume that the absorption is dominated by the X-ray line
near $5.18\,$keV and the specific opacities of the commonly used
materials are $79.4$, $0.36$ and $2.24\,$m$^2$/kg for xenon, beryllium and
polyimide, respectively. In our experiments we also use acrylic and gold,
but these materials are outside the view of the radiograph, so that we do
not need their specific opacity values accurately. For acrylic we use the
same value as for polyimide and we use the large value $10^{10}\,$m$^2$/kg for
gold.

The parallel algorithm is implemented in the following way: For each
ray and each grid block we determine first the segment of the ray if it
intersects the block. Then we integrate for each ray and block the
optical depth \ref{depth} using a trapezoidal rule and a tri-linear
interpolation. The step size of the integration is proportional to the cell
size of the block. Once the integration is done for all blocks, we add up for
each ray all integrals over the block segments using a call to
MPI\_reduce. For the $rz$-geometry we consider the 2D blocks as rings
in 3D with rectangular cross-section. The integration along the ray segments
is performed in 3D, but we use a bi-linear interpolation to obtain $\rho\kappa$
at the required integration points.

\begin{figure}
{\resizebox{\textwidth}{!}{\includegraphics[clip=]{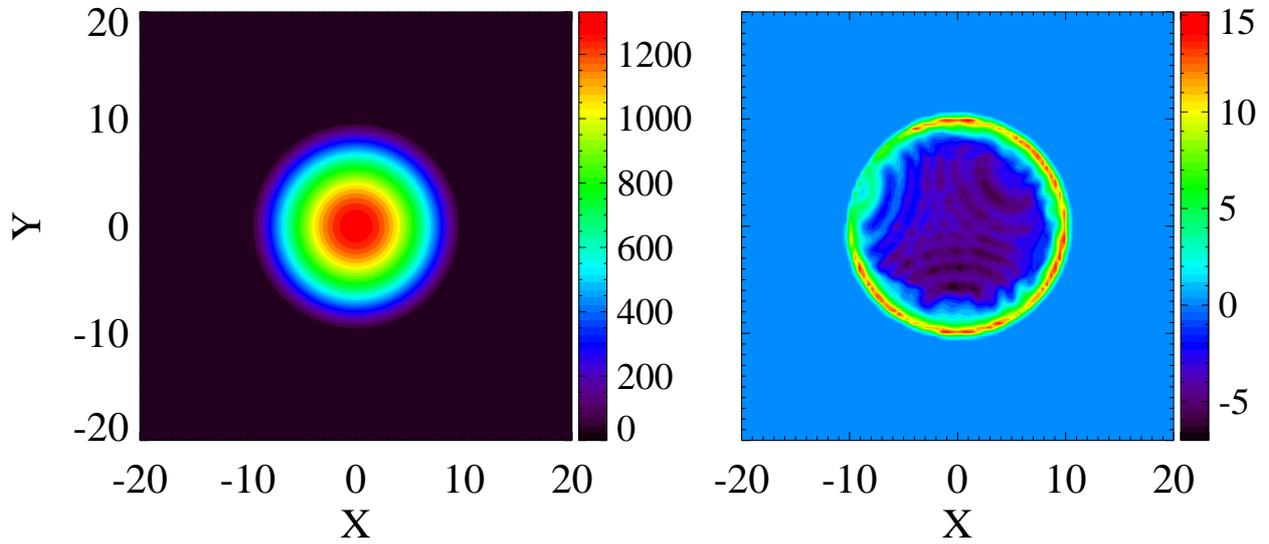}}}
\begin{center}
\end{center}
\caption{The radiograph image for an analytical density sphere on a
non-uniform mesh and a specific opacity with value one. The X-ray source is
located at $(x,y,z)=(60,80,100)$. The left panel shows the simulated image,
while the right panel is for the error relative to the analytical solution.}
\label{fig:los}
\end{figure}

\begin{figure}
{\resizebox{0.48\textwidth}{!}{\includegraphics[clip=]{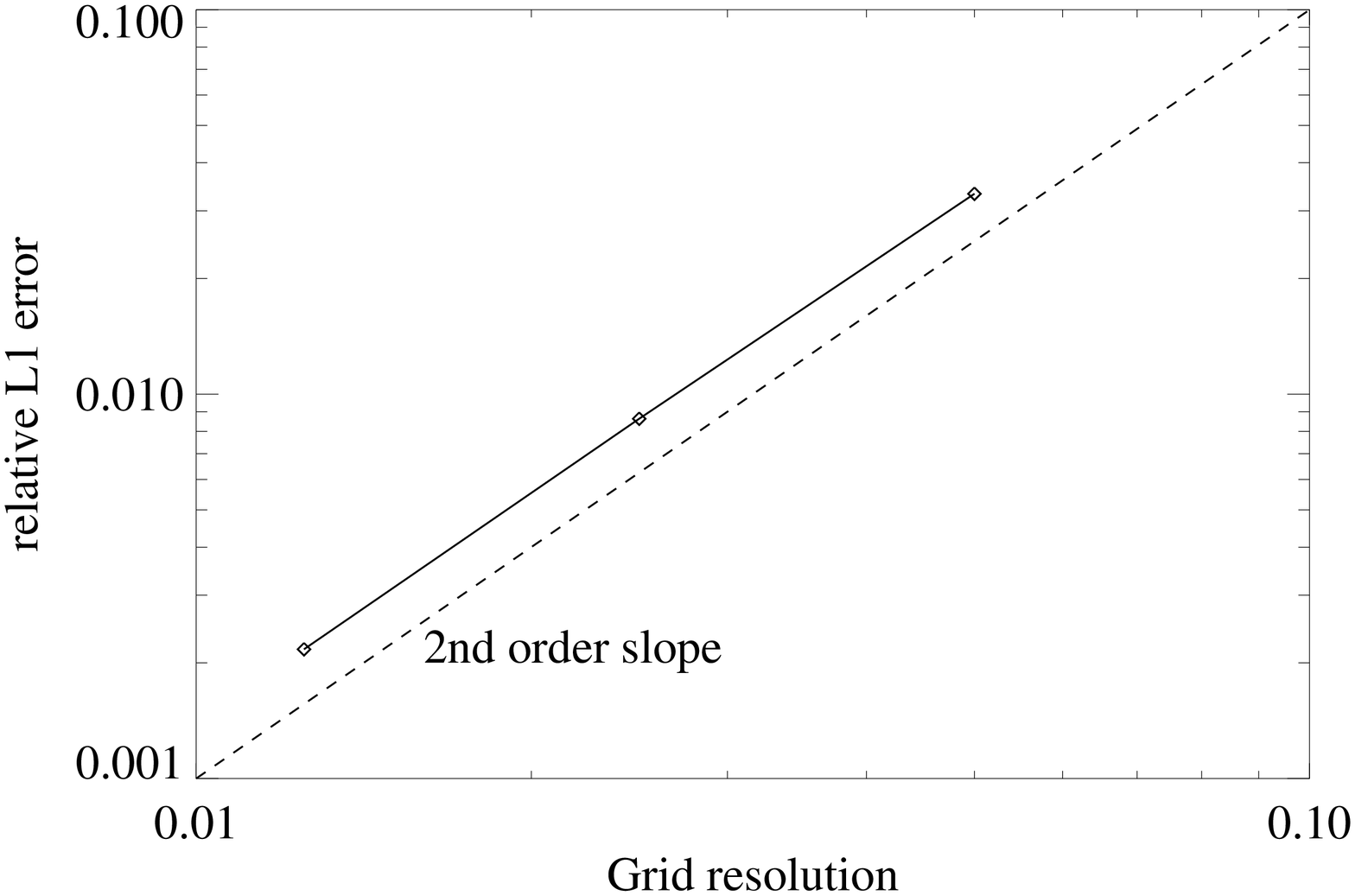}}}
\begin{center}
\end{center}
\caption{The L1 error of the optical depth relative to the analytical
reference for $20^3$, $40^3$ and $80^3$ base resolutions and one level of
refinement for $x>0$, $y>0$ and $z>0$.}
\label{fig:los_error}
\end{figure}

We have verified the implementation with an analytical problem where we
integrate along rays through a 3D sphere of a given density profile and
specific
opacity of value one. The density inside the sphere of radius $R$ is defined as
$\rho = R^2 - r^2$ if the radial distance from the center of the sphere
is $r\leq R$, while $\rho =0$ for $r>R$. The density can be integrated
analytically along a ray passing through this sphere at a distance
$\delta$ from the center. The total length of the segment of this ray
inside the sphere is $2S = 2 \sqrt{R^2 - \delta^2}$. With these definitions,
the optical depth \ref{depth} along this line segment becomes
\begin{equation}
  D = \int_{-\infty}^{\infty} \rho(s) ds
  = \int_{-S}^{S} (R^2 - \delta^2 - s^2) ds
  = \frac{4}{3} (R^2 - \delta^2)^{3/2},
\end{equation}
where $s$ is the coordinate along the ray such that $|s|=\sqrt{ r^2-\delta^2}$
is the distance to the middle of the line segment.
The radiograph image is always a plane through the origin. A pixel with
coordinates $(x',y')$ of this image is then at a distance $d=\sqrt{x'^2+y'^2}$
from the origin (here and in the following the prime indicates the coordinate
system associated with the radiograph image). The location of the X-ray
source is assumed to be on the $z'$-axis at a distance $L$ from the center of
the sphere. The ray
connecting this source and the pixel $(x',y')$ will then have a minimum
distance to the center of the sphere of $\delta = dL/\sqrt{L^2+d^2}$, as can
be deduced from similar triangles. The final formula for the optical depth
along this ray now becomes
\begin{equation}
  D = \max\left( 0, \frac{4}{3}
  \left[R^2 - \frac{L^2}{1 + L^2/(x'^2+y'^2)}\right]^{3/2} \right).
  \label{eq:exact}
\end{equation}
In the verification test we place the X-ray source at $(x,y,z)=(60,80,100)$
and the radiograph image is always orthogonal to the line connecting the
source and center of the sphere, so that the distance between the source and
image is $L=100\sqrt{2}$.  The density sphere has a radius
$R=10$. The left panel in Fig. \ref{fig:los} shows the simulated image for
a grid resolution of $40^3$ cells at the base level and one level of refinement
for $y>0$, $z>0$. The right panel shows the difference relative to
the analytical formula \ref{eq:exact}. Note that the error is smaller where
the rays go through the refined region of the grid. The grid convergence is
seen in Fig. \ref{fig:los_error} for three different
resolution at the base level: $20^3$, $40^3$, and $80^3$.
The relative error is calculated from
\begin{equation}
  E_{L1} = \frac{\sum_{i=1}^I |D_i-D_{{\rm ref}i}|}
  {\sum_{i=1}^I |D_{{\rm ref}i}|},
\end{equation}
where $D$ is the simulated optical depth, $D_{\rm ref}$ is the analytical
reference \ref{eq:exact} and $i=1,\ldots I$ indexes the pixels of the image.
We obtain second-order convergence. We have also verified the implementation
for the $rz$-geometry, in which case the sphere is a circle rotated around
the symmetry axis.

\begin{figure}
{\resizebox{0.48\textwidth}{!}{\includegraphics[clip=]{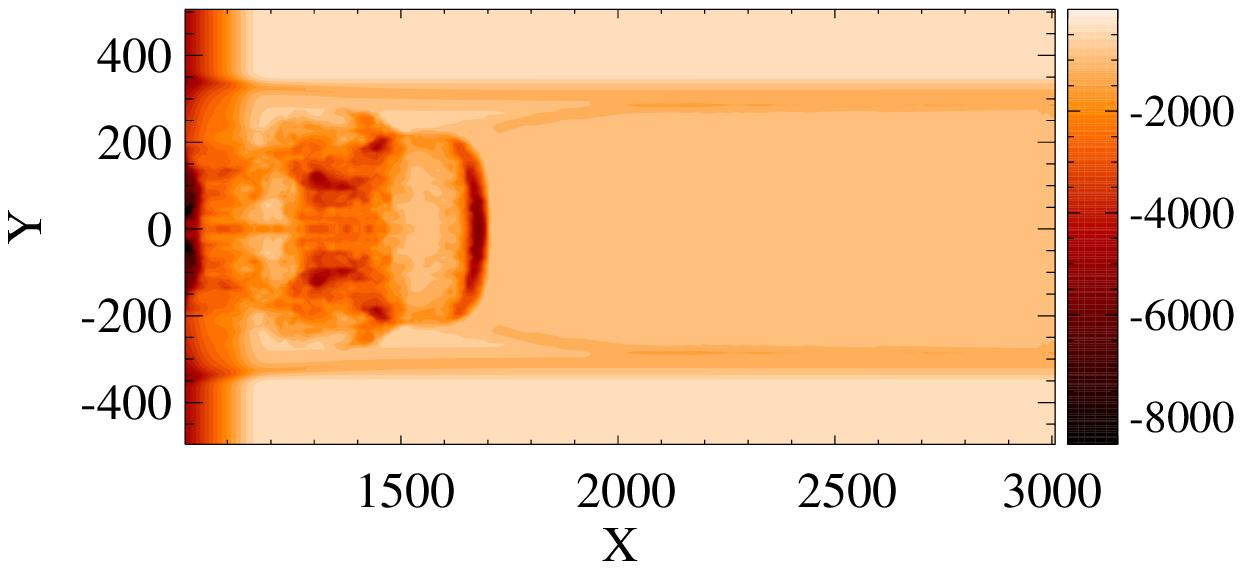}}}
{\resizebox{0.48\textwidth}{!}{\includegraphics[clip=]{circular_radiograph.eps}}}
\begin{center}
\end{center}
\caption{Original synthetic radiograph (left panel) and the processed image
after smoothing and adding Poisson noise to mimic the experimental radiograph
(right panel).}
\label{fig:radiograph}
\end{figure}

The simulated radiographs can be compared with experimental backlit pinhole
radiographs for validation studies. To make these images appear more similar,
we
need to take into account the finite pinhole size of about $20\,$mm diameter,
the finite exposure time of about $0.2\,$ns and the effect of a finite number
of collected photons (typically $50$ photons per $100\,$nm$^2$ image
pixel). The first two effects are approximated by smoothing the synthetic
radiograph over a few pixels. The finite photon count can be taken into
account by using
\begin{equation}
  D' = -\ln[ P(50) e^{-D}] = D - \ln P(50),
\end{equation}
instead of D, where $P(50)$ is a random number with a Poisson distribution and
a mean value of 50. The second equality shows that this is the same as
subtracting the logarithm of these random numbers from the original simulated
radiograph during the post-processing step. In Fig. \ref{fig:radiograph} the
original synthetic radiograph is compared with the post-processed image.

\section{Improved diffusion operator at resolution changes}
\label{sec:reschange}

In this appendix we present an improved conservative and spatially second-order
implicit scheme for the radiation diffusion and heat conduction. For
convenience, we only derive this scheme for the heat conduction. The
generalization for radiation diffusion is straightforward.

Discretizing the electron thermal heat conduction implicitly in time leads to
the linearized backward Euler equation for the electron temperature $T$
\begin{equation}
  C_{Ve}^*\frac{T^{n+1}-T^*}{\Delta t} = \nabla\cdot C_e^*\nabla T^{n+1},
\end{equation}
where $C_{Ve}$ is the electron specific heat and $C_e$ is the heat conduction
coeffcient. The time level $*$ corresponds to the state before the implicit
update. During the implicit advance with time step $\Delta t$, the coefficients
are frozen in at time level
$*$, resulting in a temporally first order scheme in general. This equation
can be recast in a linearized equation for the change $\Delta T=
T^{n+1}-T^*$:
\begin{equation}
  \left[ \frac{C_{Ve}^*}{\Delta t} - \nabla\cdot C_e^*\nabla \right] \Delta T
  = \nabla\cdot C_e^*\nabla T^*. \label{eq:implicitheatcond}
\end{equation}
The right-hand-side depends only on time level $*$. Once Eq.
\ref{eq:implicitheatcond} is solved using a linear solver, the electron
energy density can be updated using
\begin{equation}
  E_e^{n+1} = E_e^* + C_{Ve}^* (T^{n+1} - T^*),
\end{equation}
to conserve the energy.

A discrete set of equations is obtained by applying a finite volume method to
Eq. \ref{eq:implicitheatcond}. To make this scheme spatially second-order
accurate on a uniform mesh, we need a second order accurate thermal heat flux
at the face centers. This is achieved by approximating the gradient of the
electron temperature with a central difference using the cell-centered
values
\begin{equation}
  - \int_{V_i} \nabla\cdot(C_e\nabla T)dV
  = \sum_{j} S_{ij}C_{eij}\frac{T_{i} - T_{j}}{|{\bf x}_i - {\bf x}_j|},
  \label{eq:heatflux}
\end{equation}
where the control volumes are indexed by $i$, each having a volume $V_i$.
The index $j$ is for the neighboring cells having a common interface with area
$S_{ij}$ and a distance between the cell centers is $|{\bf x}_i - {\bf x}_j|$.
The heat conduction coefficient at the face is the arithmatic average of the
coefficient at the two neighboring cell centers, $C_{eij}=(C_{ei}+C_{ej})/2$.

\begin{figure}
{\resizebox{0.35\textwidth}{!}{\includegraphics[clip=]{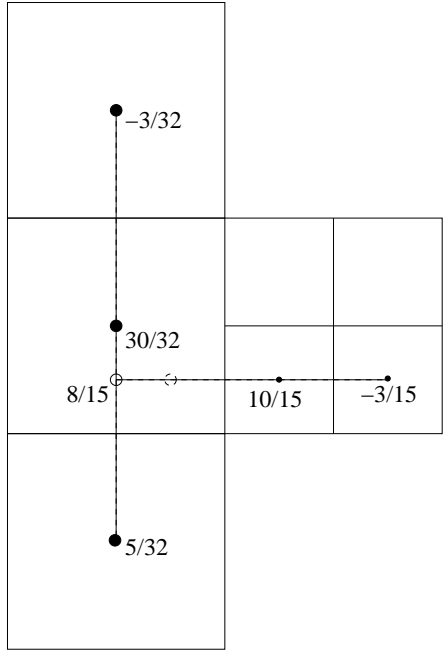}}}
\begin{center}
\end{center}
\caption{The coefficients of the third-order interpolation near the resolution
changes (similar to \cite{toth2008}). The order of the interpolation is first
along the coarse grid cells tangential to the resolution change, followed by
an interpolation along the fine cells to obtain a third-order temperature
value in the fine ghost cell indicated by the dashed circle.}
\label{fig:grid}
\end{figure}

To obtain a second-order heat flux at the resolution change, we need a
third-order interpolation of the temperature in the ghost cells. Such an
interpolation was previously used in the context of Hall magnetohydrodynamics
(MHD) \cite{toth2008}. This interpolation is only needed for the fine cells,
since the flux at the coarse side will be obtained as the sum of the fluxes at
the neighboring fine cells to preserve conservation of the scheme
\cite{berger1989}. The implementation for the heat conduction is
different from the Hall MHD, since for the heat conduction we also have to
maintain positivity of the temperature and avoid spurious oscillations. For
convenience we will restrict the
analysis to two-dimensional domains. The third-order interpolation for the
temperature value at the fine ghost cell $(0,j)$, indicated by the dashed
circle, is first performed along the
coarse cell values in the transverse direction to obtain the temperature at
$(-1/2,j)$ as depicted in Fig. \ref{fig:grid}:
\begin{equation}
  T_{-1/2,j} = \frac{5T_{-1/2,j-3/2}+30T_{-1/2,j+1/2}-3T_{-1/2,j+5/2}}{32}.
\end{equation}
To guarantee the positivity of the interpolated temperature, the value is
clipped by the maximum and minimum values of the surrounding points
\begin{eqnarray}
  T_{-1/2,j} = \max\{\min[T_{-1/2,j},
    &&\max(T_{-1/2,j-3/2},T_{-1/2,j+1/2})] , \\
  &&\min(T_{-1/2,j-3/2},T_{-1/2,j+1/2})\}.
\end{eqnarray}
The value in the fine ghost cell can now be obtained by a parabolic
interpolation along the fine cells in the direction normal to the refinement
interface
\begin{equation}
  T_{0,j} = \frac{8T_{-1/2,j}+10T_{1,j}-3T_{2,j}}{15}.
\end{equation}
For this interpolation we need again to clip the obtained value with the
surrounding temperatures
\begin{equation}
  T_{0,j} = \max\{\min[T_{0,j},\max(T_{-1/2,j},T_{1,j})],
  \min(T_{-1/2,j},T_{1,j})\}.
\end{equation}
This ghost cell value is used in the heat conduction formula
\ref{eq:heatflux}. We still need a second order accurate heat conduction
coefficient $C_{eij}$ at the fine face center. This amounts to a second-order
prolongation of the heat conduction coefficient to obtain the ghost-cell
values at the fine side, followed by an averaging of the cell centered
coefficients to the
face center. Once the thermal heat fluxes at the fine side are obtained,
conservation is restored by copying the fine fluxes to the coarse side at the
resolution changes. Note that this scheme is different from \cite{edwards1996}
where conservation of the flux at the resolution changes is enforced in the
strong sense by enforcing the flux on the coarse side to be equal to each
of the fluxes on the fine side.

In the analysis we assumed a Cartesian mesh. Generalization to curvilinear
grids is presented in \cite{toth2008}. The same third-order interpolation
procedure with clipping is also used for heat conduction along the magnetic
field lines in the context of solar wind modeling \cite{vanderholst2010}.

To verify that the improved implicit heat conduction and radiation solver
is second order, we first demonstrate a uniform heat conduction test in
$rz$-geometry. This test follows the time evolution of the electron
temperature profile for a purely heat conductive plasma similar as described
in \cite{vanderholst2011}.
We set the electron specific heat to be one and further assume the heat
conductivity $C_e$ to be a constant. The time evolution for this problem is a
product of a Gaussian profile in the $z$-direction and the Bessel function
$J_0$ in the $r$-direction:
\begin{equation}
  T = T_{\rm min} + T_0 \frac{1}{\sqrt{4\pi C_e t}}
  e^{-\frac{z^2}{4C_e t}} J_0(b r)e^{-b^2 C_e t}, \label{eq:temperature}
\end{equation}
where $b\approx3.8317$, $T_{\rm min}=3$, $T_0=10$, and $C_e=0.1$.

The domain size for this problem is $-3<z<3$ and $0<r<1$, which we decompose
in $3\times 3$ grid blocks of $30\times 30$ cells each. The central block,
$-1<z<1$ and $1/3<r<2/3$, is mesh refined by one level. At the outer boundary
we fix the ghost cells to the exact solution \ref{eq:temperature}, except for
$r=0$, in which case we apply a symmetry condition. We simulate the
time evolution from $t=1$ to $t=3/2$ using the GMRES iterative solver with a
Schwarz-type ILU preconditioner. To achieve
second-order time integration we use the Crank-Nicolson approach. The latter is
possible since the coefficients in the problem are all temporally invariant.

\begin{figure}
{\resizebox{0.48\textwidth}{!}{\includegraphics[clip=]{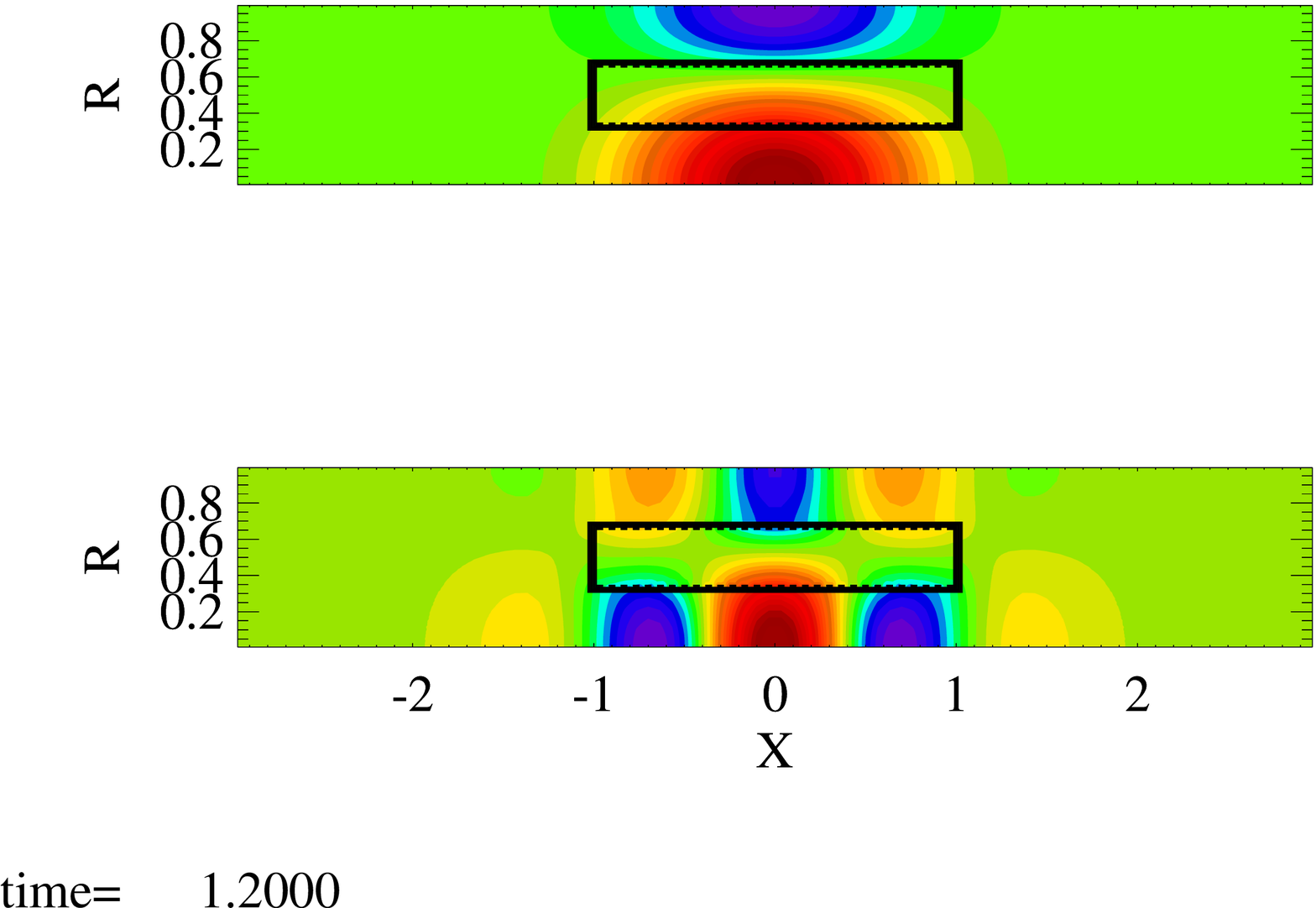}}} \\
{\resizebox{0.48\textwidth}{!}{\includegraphics[clip=]{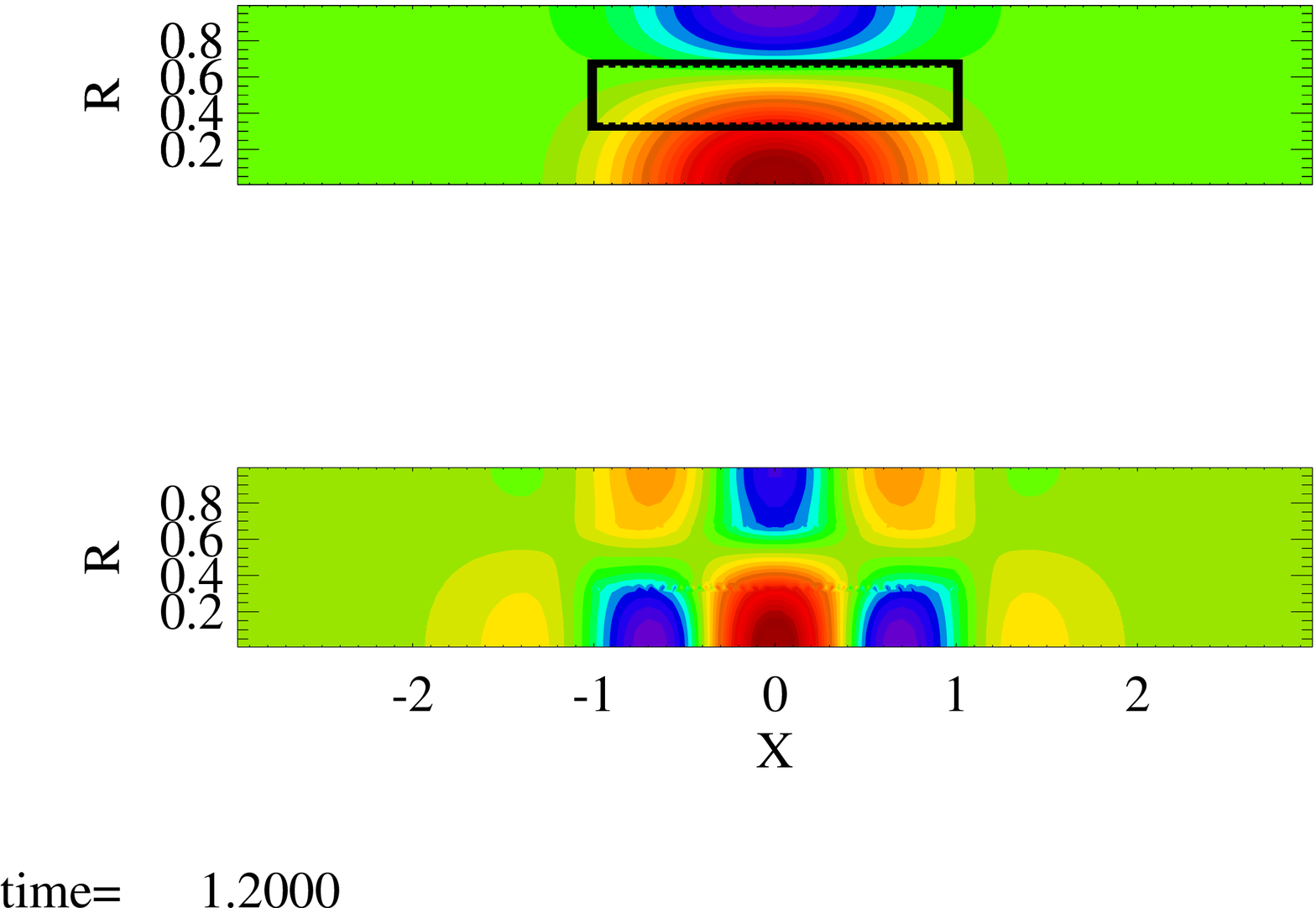}}} \\
{\resizebox{0.48\textwidth}{!}{\includegraphics[clip=]{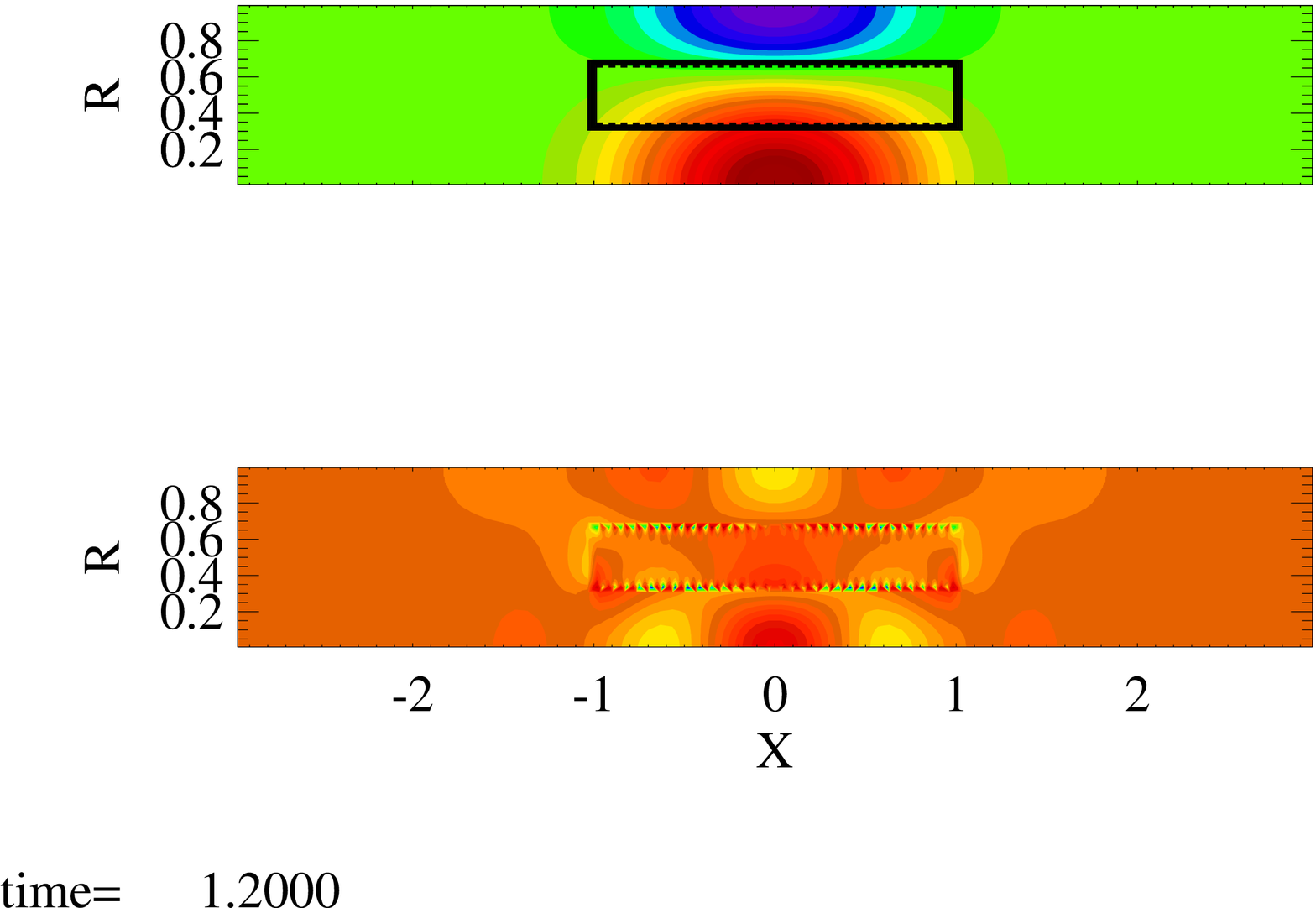}}}
\begin{center}
\end{center}
\caption{Heat conduction test in $rz$-geometry with uniform coefficients.
The top panel shows the electron temperature in color contour at the final
time $t=3/2$. The resolution changes are indicated by the black box.
The middle panel is for the temperature error relative to the reference solution, while the bottom panel shows the relative error of the original heat
conduction scheme \cite{vanderholst2011}.}
\label{fig:uniform}
\end{figure}

The final solution with $90^2$ base resolution is shown in the top panel of
Fig. \ref{fig:uniform}. The color is for the electron temperature. The
change in the grid resolution with one level of refinement is indicated by
the black line. To demonstrate that the temperature error at these resolution
changes is not significantly larger than at the uniform part of the mesh,
we also plot the spatial distribution of the temperature error in the bottom
panel. The maximum error is clearly at the coarse grid away from the resolution
changes.

\begin{figure}
{\resizebox{0.48\textwidth}{!}{\includegraphics[clip=]{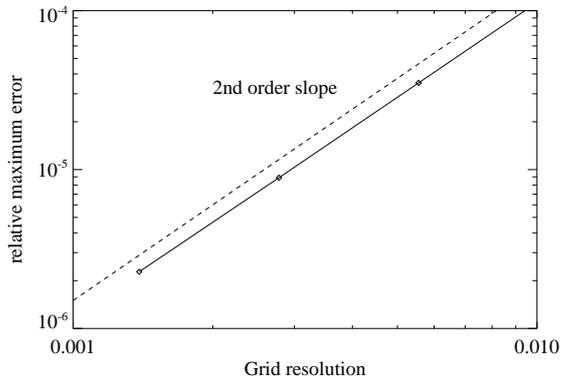}}}
\begin{center}
\end{center}
\caption{Relative maximum error for the heat conduction test in $rz$-geometry
and with uniform coefficients. Second-order convergence is achieved on a
non-uniform grid.}
\label{fig:uniform_error}
\end{figure}

We use the relative maximum error to study the grid convergence. This error is
defined by
\begin{equation}
  E_{\mbox{L}\infty} = \frac{\max_{i=1,\ldots,I}
    |{T}_{i}-{T_{\rm ref}}_{i}|}{\max_{i=1,\ldots,I}|
    {T_{\rm ref}}_{i}|},
\end{equation}
in which $T$ is the numerical solution on the $i=1,\ldots,I$ control
volumes and $T_{\rm ref}$ is the analytical reference in Eq.
\ref{eq:temperature}. The second-order convergence rate is depicted in
Fig. \ref{fig:uniform_error} for grid resolutions of $90^2$, $180^2$, $360^2$,
and $720^2$ at the base level.

To demonstrate that the new heat conduction and radiation solver does not
under- or over-shoot near discontinuities, we show the Mach 5 non-equilibrium
gray-diffusion test of \cite{lowrie2008}. This test uses non-uniform
radiation diffusion coefficient and Planck opacity that depend on the density
and temperature as defined by $D_r=0.0175(\gamma T)^{7/2}/\rho$ and
$c\kappa_P = 10^6/D_r$. In \cite{vanderholst2011}, this verification was
transformed to a heat conduction test. Here, we will keep it as a radiation
test. We add a Mach -5 flow to the 1D initial condition,
so that both the shock and radiation condition will move to the left
with a Mach -5 velocity. This solution is then rotated counter-clockwise
over an angle $\tan^{-1}(1/2)$ on a 2D grid to make the test problem more
difficult. The domain size is $-0.0384<x<0.0384$ by $-0.0048<y<0.0048$ and
within the region $-0.0128<x<0.0128$ and $-0.0016<y<0.0016$ we refine the
mesh by one level. The grid is designed in such a way that during the time
evolution both the shock and the radiation precursor front travel through
a resolution change. For the simulation of the hydrodynamic part, we use
the HLLE scheme with the generalized Koren limiter by setting $\beta=3/2$
and a CFL of 0.8. The radiation diffusion and energy exchange between the
radiation and material are solved implicitly with the new implicit method
using the GMRES iterative solver with a block ILU preconditioner.

\begin{figure}
\begin{center}
{\resizebox{0.48\textwidth}{!}{\includegraphics[clip=]{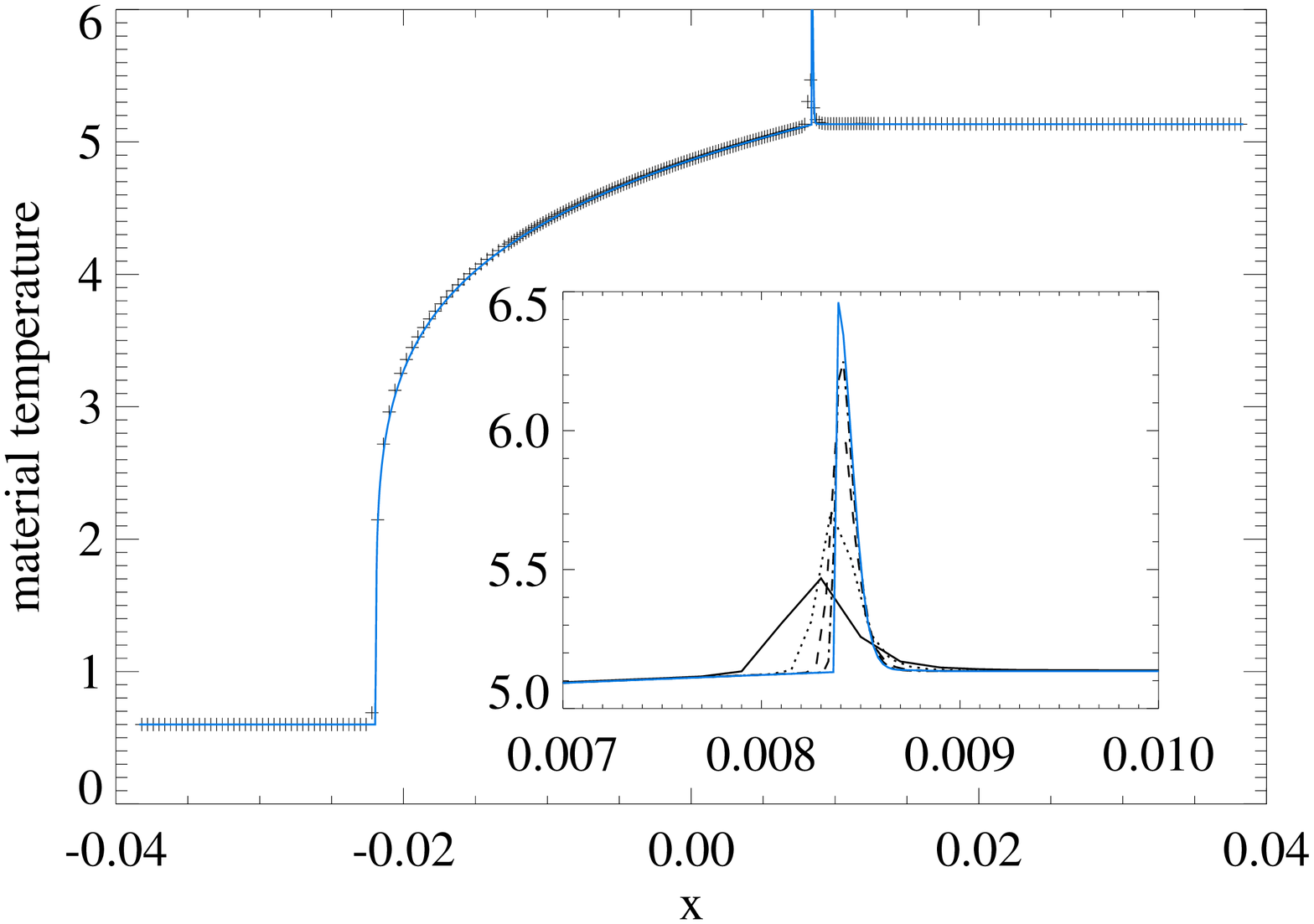}}}
{\resizebox{0.48\textwidth}{!}{\includegraphics[clip=]{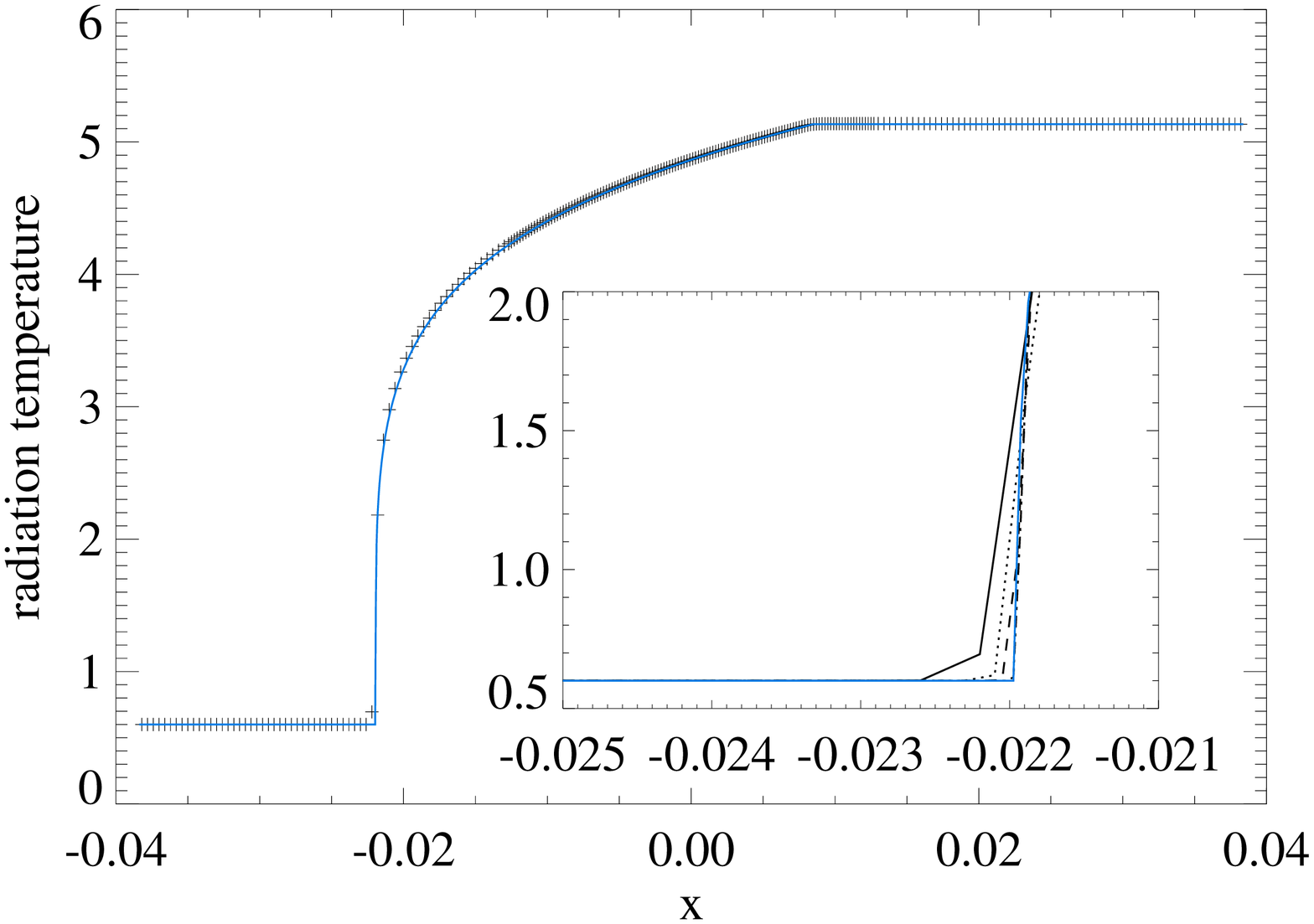}}}
\end{center}
\caption{Material (left panel) and radiation (right panel) temperature at
the final time for the Mach 5 radiative shock tube problem rotated on a
non-uniform grid. The blue line is the reference solution obtained from
\cite{lowrie2008}. The insets show the temperature jump at the hydro-shock
(left panel) and the radiation precursor front (right panel).}
\label{fig:lowrie3}
\end{figure}

The material and radiation temperatures at the final time, $t=0.0025$,
are shown in the left and right panel of Fig. \ref{fig:lowrie3}, respectively.
The solid, dotted, dashed, and dot-dashed black lines indicate the
base resolutions of $192\times 24$, $384\times 48$, $768\times 96$,
and $1536\times 192$, respectively. The blue line is the semi-analytic
solution of \cite{lowrie2008}. The new scheme correctly solves the precursor
front at $x\approx -0.022$ and shock front at $x\approx 0.0085$ without
spurious oscillations.

\bibliographystyle{elsarticle-num}

\end{document}